\pdfoutput=0
\documentclass[useAMS,usenatbib,usegraphicx]{mn2e}
\usepackage{times}
\usepackage{color}
\usepackage{aas_macros}

\title[E/B-mode Estimation with Binned Cosmic Shear Data]{Cosmic Shear E/B-mode Estimation with Binned Correlation Function Data}
\author[M. R. Becker]{Matthew R. Becker\thanks{E-mail: beckermr@uchicago.edu}\\
Department of Physics, 5720 S. Ellis Avenue, The University of Chicago, Chicago, IL 60637\\
Kavli Institute for Cosmological Physics, 5640 South Ellis Avenue, The University of Chicago, Chicago, IL 60637}

\begin{document}

\date{}

\pagerange{\pageref{firstpage}--\pageref{lastpage}} \pubyear{2012}

\maketitle

\label{firstpage}

\begin{abstract}
In this work I study the problem of E/B-mode separation with binned cosmic shear two-point correlation function data.  Motivated by previous work on 
E/B-mode separation with shear two-point correlation functions and the practical considerations of data analysis, I consider E/B-mode estimators which are linear 
combinations of the binned shear correlation function data points.  I demonstrate that these estimators mix E- and B-modes generally.  
I then show how to define estimators which minimize this E/B-mode mixing and give practical recipes for their construction and use.  Using these optimal estimators, I 
demonstrate that the vector space composed of the binned shear correlation function data points can be decomposed into approximately ambiguous, E- and B-mode subspaces.  
With simple Fisher information estimates, I show that a non-trivial amount of information on typical cosmological parameters is contained in the ambiguous mode subspace 
computed in this formalism.  Next, I give two examples which apply these practical estimators and recipes to generic problems in cosmic shear data analysis: data compression 
and spatially locating B-mode contamination.  In particular, by using wavelet-like estimators with the shear correlation functions directly, one can pinpoint B-mode contamination to 
specific angular scales and extract information on its shape. Finally, I discuss how these estimators can be used as part of blinded or closed-box cosmic shear data analyses in order to assess 
and find B-mode contamination at high-precision while avoiding observer biases.
\end{abstract}



\section{Introduction}\label{sec:intro}
\setcounter{footnote}{0}

Cosmic shear, or the weak gravitational lensing of background galaxies by cosmological density fields, is one of the most 
important techniques for probing the properties of Dark Energy (see, e.g., \citeauthor{weinberg2012} \citeyear{weinberg2012} for a recent review) and also 
the growth of structure predicted by General Relativity (GR) or its possible modifications \citep[e.g.,][]{schmidt2008,beynon2010,vanderveld2011}.  
Cosmic shear measurements can also help constrain other cosmologically interesting signals, such as primordial non-Gaussianity \citep[e.g.,][]{fedeli2010,marian2011,maturi2011,giannantonio2012,hilbert2012}, 
the properties of various hot and warm dark matter models \citep[e.g.,][]{schaefer2008,debono2010,markovic2011}, or the properties of neutrinos 
\citep[e.g.,][]{cooray1999,song2004,hannestad2006,kitching2008,ichiki2009,debarnardis2009,jimenez2010}.
To this end, ongoing and planned wide-field optical surveys, such as the DES\footnote{The Dark Energy Survey - http://www.darkenergysurvey.org}, 
LSST\footnote{Large Synoptic Survey Telescope - http://www.lsst.org}, Euclid\footnote{http://sci.esa.int/euclid}, 
WFIRST\footnote{Wide-Field Infrared Survey Telescope - http://wfirst.gsfc.nasa.gov}, 
HSC\footnote{Hyper Suprime-Cam - http://www.naoj.org/Projects/HSC}, KIDS\footnote{The Kilo Degree Survey - http://kids.strw.leidenuniv.nl}, and 
Pan-STARRS\footnote{The Panoramic Survey Telescope \& Rapid Response System - http://pan-starrs.ifa.hawaii.edu} surveys, 
will measure the shapes of hundreds of millions to billions of galaxies and thus cosmic shear signals with unprecedented statistical precision.  

Given this incredible statistical power, understanding and mitigating potential systematic errors in these measurements will be very important.  Systematic 
contamination to cosmic shear signals can arise from a variety of sources, including the process of observing and estimating galaxy shapes from pixelated images 
\citep[e.g.,][]{kaiser2000,bernstein2002,vale2004,hoekstra2004,guzik2005,paulinhenriksson2008,cypriano2010,voigt2010,kacprzak2012,refregier2012,voigt2012,antonik2012} 
or estimating photometric redshifts \citep[e.g.,][]{ma2006,huterer2006,bridle2007,sun2009,hearin2010,bernstein2010,cunha2012}.  
There are also astrophysical sources of systematic errors, like intrinsic alignments 
\citep[e.g.,][]{heavens2000,croft2000,catelan2001,crittenden2001,crittenden2002,jing2002b,lee2002,hirata2004,heymans2006b,hui2008,semboloni2008}, 
source galaxy clustering \citep{schneider2002}, or the effects of baryons and galaxy formation on the matter power spectrum 
\citep[e.g.,][]{white2004,zhan2004,huterer2005,jing2006,rudd2008,guillet2010,vandaalen2011,casarini2011,casarini2012,hearin2012}.  
These systematic errors, if left uncontrolled, can bias and/or degrade constraints on the properties of Dark Energy 
or modifications to GR from future surveys 
\citep[e.g.,][]{hirata2003,hirata2004,guzik2005,huterer2005,huterer2006,mandelbaum2006b,ma2006,bridle2007,hirata2007,hearin2009,sun2009,semboloni2009,bernstein2010,hearin2010,kirk2011,kirk2012,cunha2012,hearin2012}.

Besides direct image and structure formation simulations to study cosmic shear data analysis, systematics, and theoretical modeling in detail 
(e.g., STEP1 \citep{heymans2006}; STEP2 \citep{massey2007}; GREAT08 \citep{bridle2010}; GREAT10 \citep{kitching2012}; 
\citealp{jain2000,vale2003,lee2008,hilbert2009,sato2009,teyssier2009,hahn2010,kiessling2011,harnoisderaps2012}), it is important to distinguish between 
observational signals which can arise from GR and those which cannot \citep{kaiser1992}.  At first order in the gravitational potential, GR will only produce cosmic shear 
patterns known as E-modes (see, e.g., \citeauthor{dodelson2003} \citeyear{dodelson2003} for a pedagogical introduction).  
The complementary patterns, know as B-modes, are not produced by GR at first order, though they can be produced in small amounts at higher order 
\citep[e.g.,][]{jain2000,cooray2002,vale2003,hilbert2009,bernardeau2010,krause2010}.  Many of the sources of systematic 
contamination, though not all, can produce B-modes in addition to E-modes \citep[e.g.,][]{crittenden2001,crittenden2002,schneider2002,vale2004,hirata2004,jarvis2004b,guzik2005,antonik2012}.  
Therefore assessing B-mode contamination in cosmic shear signals can test for systematic errors throughout the various steps of cosmic shear data analysis, from observing the 
galaxies with a telescope and imaging camera, all the way through to the theoretical modeling and constraints on cosmological parameters.

Methods for separating E- and B-modes in cosmic shear data have been studied extensively by previous authors.  
Broadly, these methods either operate directly on the shear field \citep{schneider1998,seljak1998,hu2001b,heavens2003,leonard2012} or on the shear 
two-point correlation functions \citep{crittenden2002,schneider2002,schneider2007,schneider2010,fu2010}.  
E/B-mode separation in the context of higher-order correlation functions has been studied as well \citep{jarvis2004,schneider2005,shi2011,krause2012}. 
Additionally, techniques originally designed for the analysis of Cosmic Microwave Background polarization signals \citep[e.g.,][]{wandelt2001,smith2006} can also be applied to 
cosmic shear \citep{hikage2011}.  Importantly, the details of the implementation of these methods can effect their performance significantly \citep[e.g.,][]{smith2006,kilbinger2006}.  
For the shear two-point correlation functions, $\xi_{\pm}(\theta)$ defined below, \citet{schneider2007} have shown that a broad class of E/B-mode statistics, can be written in the following form
\begin{eqnarray}
E_{c} & = & \frac{1}{2}\int_{L}^{H}d\theta\,\theta\big[T_{+}(\theta)\xi_{+}(\theta) + T_{-}(\theta)\xi_{-}(\theta)\big]\nonumber\\
B_{c} & = & \frac{1}{2}\int_{L}^{H}d\theta\,\theta\big[T_{+}(\theta)\xi_{+}(\theta) - T_{-}(\theta)\xi_{-}(\theta)\big]\nonumber\ .
\end{eqnarray}
By choosing the range of integration $[L,H]$ and the forms of $T_{\pm}(\theta)$ properly, one can show that $E_{c}$ will contain only E-mode information 
and $B_{c}$ will contain only B-mode information either over an infinite interval or finite interval with $L >0$ \citep{schneider2007}.  Note that these statistics 
assume one has continuous shear correlation function data.  

Practical implementations of the $E_c$ and $B_c$ statistics in cosmic shear data analysis are constrained in several ways.  The shear correlation functions are usually 
estimated in bins of angle, say $N$ bins, with some effective binning weight $W_{i}(\theta)$.  In particular, the expectation value of the estimated shear 
correlation function data point for the $i$th bin is \citep[e.g.,][]{schmidt2009a}
\begin{equation}\label{eqn:xpmdef}
\left<\widehat{\xi}_{\pm i}\right> = \int_{L_{i}}^{H_{i}}d\theta\,W_{i}(\theta)\xi_{\pm}(\theta)\ .
\end{equation}
The form of these binning functions and this result is discussed below and in Appendix~\ref{app:srcclust}.  Thus the statistics $E_{c}$ and $B_{c}$ in this case must be 
estimated from the binned shear correlation function data, $\widehat{\xi}_{\pm i}$.  Also, in order for the statistics $E_{c}$ and $B_{c}$ to remain 
pure two-point statistics, any procedure for estimating them from cosmic shear data can only consider linear combinations of the binned shear correlation function data,
\begin{equation}\label{eqn:xpmlinstat}
X_{\pm}=\frac{1}{2}\sum_{i}\left[F_{+i}\widehat{\xi}_{+i} \pm F_{-i}\widehat{\xi}_{-i}\right]\ ,
\end{equation}
where the $F_{\pm i}$ are constants which describe the statistics \citep[see e.g.,][]{kilbinger2006b}.  

In this work, I study the use of these linear combinations for E/B-mode separation and thus the effects of these constraints on cosmic shear data analysis.  
In particular, after covering the basics of cosmic shear in Section~\ref{sec:cosmodefs}, I demonstrate in Section~\ref{sec:ebmix} that \textit{even if the statistics $E_{c}$ 
and $B_{c}$ contain pure E- and B-mode information, the statistics $X_{\pm}$ generally exhibit E/B-mode mixing due to the binning}.  In Section~\ref{sec:buildEBest}, I show 
how to define the statistics $X_{\pm}$ through the choice of the $F_{\pm i}$, so that they suppress the E/B-mode mixing below detectable levels for any current or upcoming 
cosmic shear survey.  In this section I also give practical recipes for computing the $F_{\pm i}$.  Other potential and ultimately equivalent optimal estimator definitions are discussed in 
Section~\ref{sec:otherest}.  I then show how to use these optimal estimators to divide the vector space of correlation function data points up into approximately ambiguous, E- and B-mode subspaces in Section~\ref{sec:ebmodedec}.  I compute the Gaussian covariances of these statistics in Section~\ref{sec:ebcov}.  In Sections~\ref{sec:cosebi} and \ref{sec:wvstats}, I provide two examples 
of these statistics to illustrate how to build and use them in practice.  I provide an example of how to decompose the shear correlation functions into ambiguous, E- and B-modes and also 
discuss the Fisher information content of these subspaces in Section~\ref{sec:fishinfo}. I find that the ambiguous mode subspace has a non-trivial amount of information about typical cosmological 
parameters.  Finally, I conclude and discuss how these statistics are applicable to blinded or closed-box cosmic shear data analyses in Section~\ref{sec:conc}.

\section{Cosmological Weak Lensing}\label{sec:cosmodefs}
The basic equations describing weak lensing by cosmological density fields are covered in detail in other works \citep[see
e.g.,][]{bartelmann2001,dodelson2003,hoekstra2008,bartelmann2010}.  The discussion presented here is merely a summary of the relevant results needed for this work. Given the 3D
density matter power spectrum $P(k,a)$ as a function of wave number $k$ and scale factor $a$, the 2D convergence power spectrum as a function of 2D
wave number $\ell$ is defined in the Limber approximation as \citep[cf.][]{hoekstra2008}
\begin{eqnarray}
C^{k}_{ij}(\ell) &=& \int_{0}^{\infty}d\chi\, \frac{W_{i}(\chi)W_{j}(\chi)}{\chi^{2}}P(\ell/\chi(z),a)\nonumber\nonumber\\
W_{i,j}(\chi) &=& \frac{3}{2}\Omega_{m}\left(\frac{H_{0}}{c}\right)^{2}\frac{\chi}{a(\chi)}\int_{\chi}^{\infty} d\chi_{s}\, \frac{n_{i,j}(\chi_{s})}{\bar{n}_{i,j}}\frac{\chi_{s}-\chi}{\chi_{s}}\nonumber
\end{eqnarray}
where $z$ is the redshift, $\chi(z)$ is the comoving distance, and $n_{i}(\chi)$ is the redshift distribution of the lensing sources for source set $i$ normalized 
to the total source density,  $\bar{n}_{i,j}=\int d\chi_{s}\,n_{i,j}(\chi_{s})$.  These expressions assume straight-line photon paths from the sources to the observer, 
commonly called the Born approximation, and a spatially flat universe.  I use the non-linear power spectrum
fitting formula of \citet{smith2003} to compute $C_{ij}^{k}$ and the fitting formula of \citet{eisenstein1998} to evaluate the linear power spectrum.
Additionally, in this work all lensing sources are at a single redshift, $z_{s}=1.0$, such that $\bar{n}(\chi_{s})=\delta(\chi_{s} - \chi(z_{s}))$,
where $\delta(\chi)$ is the Dirac delta function.  

In cosmological weak lensing, one observes the shear field (neglecting reduced shear effects, see e.g., 
\citeauthor{schneider1995a} \citeyear{schneider1995a}; \citeauthor{mandelbaum2006} \citeyear{mandelbaum2006}) along with an assumed to be random contribution from 
galaxy shapes and orientations.  This last effect is commonly called shape noise and is characterized by the shape noise per component, $\sigma_{e}$.  The breaking of the assumption of 
random galaxy orientations is generically referred to as intrinsic alignments and is a primary source of systematic error in cosmic shear measurements (see the references given above).
Given the complex shear field, $\gamma = \gamma_{1} + i\gamma_{2}$, one can define the $\xi_{\pm}$ correlation functions as
\citep[cf.][]{schneider2002}
\begin{eqnarray}
\xi_{+} &=& \left<\gamma_{t}\gamma_{t}\right> +  \left<\gamma_{\times}\gamma_{\times}\right>\nonumber\\
\xi_{-} &=& \left<\gamma_{t}\gamma_{t}\right> -  \left<\gamma_{\times}\gamma_{\times}\right>\nonumber
\end{eqnarray}
where $\gamma_{t}=-Re(\gamma e^{-2i\phi})$, $\gamma_{\times}=-Im(\gamma e^{-2i\phi})$, and $\phi$ is the polar angle of the vector connecting the 
two points.  In Appendix~\ref{app:srcclust}, I present the standard expressions for galaxy pair-wise shear 
correlation function estimators and their expectation values \citep[see, e.g.,][]{schneider2002b,schmidt2009a}.

In Fourier-space, the shear field is typically separated into a component with no net handedness, the E-mode part, and a handed component, the B-mode part. 
In terms of the power spectra of the E- and B-mode parts, these correlation functions are \citep[cf.][]{schneider2007}
\begin{eqnarray}
\xi_{+}(\theta) &=& \int_{0}^{\infty}\frac{d\ell\,\ell}{2\pi}J_{0}(\ell\theta)\left[P_{E}(\ell) + P_{B}(\ell)\right]\label{eqn:xipell}\\
\xi_{-}(\theta) &=& \int_{0}^{\infty}\frac{d\ell\,\ell}{2\pi}J_{4}(\ell\theta)\left[P_{E}(\ell) - P_{B}(\ell)\right]\label{eqn:ximell}
\end{eqnarray}
where the $J_{n}(\ell\theta)$ are cylindrical Bessel functions.  In the Born approximation, the B-mode power is identically zero, $P_{B}(\ell)\equiv0$, and the E-mode power is equal to the 
convergence power spectrum, $P_{E}(\ell)=C^{{k}}_{ij}$.  Corrections to the Born approximation are very small \citep[e.g.,][]{jain2000,cooray2002,vale2003,hilbert2009,bernardeau2010,krause2010}.  
The shape noise contribution to the power spectra, $\sigma_{e}^{2}/\bar{n}$, has been purposefully left out of these expressions 
because pair-wise estimators of the shear correlation functions (see Appendix~\ref{app:srcclust}) do not exhibit noise biases \citep{schneider2002b}.  

Finally, the covariance of the shear-shear correlation functions can be computed under the assumption that the shear fields are Gaussian with the
following expressions from \citet{joachimi2008}
\begin{eqnarray}
\lefteqn{\left<\xi_{+/-}(\theta_{1})\xi_{+/-}(\theta_{2})\right>=}\nonumber\\
&&\frac{1}{\pi\Omega_{s}}\int_{0}^{\infty}d\ell\,\ell\, J_{0/4}(\ell\theta_{1})J_{0/4}(\ell\theta_{2})\times\bigg\{P^{2}_{E}(\ell) + P^{2}_{B}(\ell)\nonumber\\
&&\ \ \ \ \ \ \ \ \ \left.+ \frac{2\sigma_{e}^{2}}{\bar{n}}\left[P_{E}(\ell) + P_{B}(\ell)\right]\right\} + \delta_{\theta_{1}\theta_{2}}\frac{4\sigma_{e}^{4}}{\bar{n}^{2}2\pi\theta_{1}\Delta\theta_{1}\Omega_{s}}\nonumber
\end{eqnarray}
\begin{eqnarray}
\lefteqn{\left<\xi_{+}(\theta_{1})\xi_{-}(\theta_{2})\right>=}\nonumber\\
&&\frac{1}{\pi\Omega_{s}}\int_{0}^{\infty}d\ell\,\ell\, J_{0}(\ell\theta_{1})J_{4}(\ell\theta_{2})\nonumber\\
&&\ \ \ \ \ \ \ \ \ \times\bigg\{P^{2}_{E}(\ell) + P^{2}_{B}(\ell) + \frac{2\sigma_{e}^{2}}{\bar{n}}\left[P_{E}(\ell) + P_{B}(\ell)\right]\bigg\}\label{eqn:shearcov}
\end{eqnarray}
where $\Delta\theta_{1}$ is the bin width and $\Omega_{s}$ is the survey area.  Note that the shape noise contributes to cross terms in braces in
addition to the diagonal terms given by the Kronecker delta function.  The covariance between $\xi_{+}$ and $\xi_{-}$ is given by second of the above
expressions.  

The expressions for the correlation function covariance matrix will be useful below for computing the Fisher information content of the shear correlation functions under the assumption the errors 
are Gaussian.  The Fisher information matrix is \citep[e.g.,][]{tegmark1997}
\begin{equation}
F_{ij} = \frac{1}{2}\mathrm{Tr}\left[\mathrm{A}_{i}\mathrm{A}_{j} + \mathrm{C}^{-1}\mathrm{M}_{ij}\right]
\end{equation}
where $\mathrm{C}$ is the covariance matrix of the observations, $\mathrm{A}_{i}=\mathrm{C}^{-1}\mathrm{C}_{,i}$, and 
$\mathrm{M}_{ij}=\vec{\mu}_{,i}\,\vec{\mu}^{T}_{,j} + \vec{\mu}_{,j}\,\vec{\mu}^{T}_{,i} $.  Here $\vec{\mu}$ is the vector of mean values of the data and the 
notation $,i$ indicates a partial derivative with respect to parameter $\theta_{i}$.  Below I neglect the information in the covariance matrix so that the Fisher information is 
computed only using the last term in the equation above.  There are several conventions in the literature for comparing the Fisher information 
content of various analyses.  I roughly follow \citet{schneider2010} and simply compare various analyses by computing $f \equiv \sqrt{|\mathrm{F}|}$ for a fiducial set of parameters 
for each analysis.  I use $\sigma_{8}$, the normalization of the linear matter power spectrum today filtered in 8 $h^{-1}$Mpc spheres, and $\Omega_{m}$, the mean matter density today in units 
of the critical density, as the fiducial set of parameters for comparison.  These Fisher information estimates are not meant to be realistic survey projections, but to give a sense of how much of the total 
Fisher information is retained by the E/B-mode statistics presented in this paper.

Throughout this work I assume a flat $\Lambda$CDM universe with $\Omega_{m}=0.25$, $H_{0}=h^{-1}100$ kms$^{-1}$Mpc$^{-1}$, $h=0.7$, 
$\sigma_{8}=0.8$, $n_{s}=1.0$, and $\Omega_{b}=0.044$.  Also, I consider two different prototypical weak lensing surveys with different areas, $\Omega_{s}$, and 
lensing source number densities, $\bar{n}$.  The first survey has $(\Omega_{s},\bar{n})=(5000\ \mathrm{deg}^{2},10\ \mathrm{gals/arcmin}^{2})$, 
typical of the DES and the second has $(\Omega_{s},\bar{n})=(20,000\ \mathrm{deg}^{2},40\ \mathrm{gals/arcmin}^{2})$, typical of the LSST survey.  
I set the lensing shape noise per component to $\sigma_{e}=0.3$ for both the DES- and LSST-like surveys.

\section{E/B-mode Estimation with Binned Data}
In this section I study the problem of E/B-mode estimation with binned cosmic shear data in detail.  I show that generally estimators which are 
linear combinations of the binned cosmic shear data points mix E- and B-mode signals.  I then demonstrate how to define optimal estimators which minimize this E/B-mode 
mixing and give practical recipes for constructing and using these estimators.  Next I discuss other potential optimal estimator definitions, demonstrating that they are approximately equivalent 
to the definition used throughout this work.  Then I discuss how to decompose the vector space composed of the binned shear correlation functions 
into approximately ambiguous, E- and B-mode subspaces. 
Finally, I compute the variance and covariance of these estimators under the assumption the shear power spectra are Gaussian.

\subsection{Binned Data \& E/B-mode Mixing}\label{sec:ebmix}
As stated above, shear correlation functions are typically measured in a bins of angle so that $\theta\in[L_{i},H_{i}]$ for the $i$th bin.  Additionally, for the standard shear two-point function estimator, 
the shear correlation function measurements are weighted by the bin weighting function $W_{i}(\theta)$, as in Equation~(\ref{eqn:xpmdef}).  Generally these bin weighting 
functions are normalized to unity so that $\int_{L_{i}}^{H_{i}}d\theta W_{i}(\theta)=1$.  Throughout this work I assume the bin weighting functions are normalized properly.  
As detailed in Appendix~\ref{app:srcclust}, for pair-wise estimators of the shear correlation functions this bin weighting function is in general quite complicated  because of the survey 
window function and source galaxy clustering.  Note however that for unclustered sources and neglecting the survey window function, $W_{i}(\theta)=2\theta/(H_{i}^{2}-L_{i}^{2})$, so 
that the dominant effect is a purely geometric weighting.  This effect arises from the increase in the number of galaxy pairs due to the increase in the area in the outer part of the bin relative 
to the inner part.  I show in Appendix~\ref{app:srcclust} that by using small enough bins, the effects of the source clustering can be made negligible.  Assessing the form and magnitude 
of the weightings from the survey window function is beyond the scope of this work.  Thus when needed, I only use the geometric weightings given above.  

Given the bin window functions, I now consider statistics which are general linear combinations of the shear correlation function data 
points described by Equation~(\ref{eqn:xpmlinstat}).  The form of these statistics has been chosen for easy 
comparison with work for E/B-mode statistics for unbinned shear correlation function data presented by \citet{schneider2007}.  It is important to realize however that I have included 
a bin size dependent weight through the normalization of the $W_{i}(\theta)$ in the definition of the statistics unlike \citet{schneider2007}. The correspondence between 
the $F_{\pm}$ coefficients, which define the statistics $X_{\pm}$, and the $\{E_{c},B_{c}\}$ statistics presented in \citet{schneider2007} is $T_{\pm}(\theta)=W_{i}(\theta)F_{\pm i}/\theta$ for each bin $i$.

I will now show that \textit{the statistics $X_{\pm}$ always mix E- and B-mode information}, barring a very special choice of the bin window functions $W_{i}(\theta)$.  
This derivation follows closely a derivation presented in \citet{schneider2007}, but accounts for the binning explicitly.  In order to 
proceed, I substitute the definitions of the shear correlation functions in terms of the E- and B-mode power spectra 
from Equations (\ref{eqn:xipell}) and (\ref{eqn:ximell}) into the definition of $X_{\pm}$ to get the following expression for the expectation value of the statistics
\begin{eqnarray}
\left<X_{\pm}\right> &=& \int_{0}^{\infty}\frac{d\ell\,\ell}{2\pi} P_{E}(\ell)\,W_{\pm}(\ell) + P_{B}(\ell)\,W_{\mp}(\ell)\label{eqn:xipmell}\\
W_{\pm}(\ell) &=& \frac{1}{2}\sum_{i}\left(F_{+i}\int_{L_{i}}^{H_{i}}d\theta\,W_{i}(\theta)\,J_{0}(\ell\theta)\right.\nonumber\\
&& \ \ \ \ \ \ \ \ \  \ \ \ \ \ \left.\pm F_{-i}\int_{L_{i}}^{H_{i}}d\theta\,W_{i}(\theta)\,J_{4}(\ell\theta)\right)\label{eqn:wpmdef}
\end{eqnarray}
These equations with $P_{B}(\ell)\equiv0$ were presented by \citet{kilbinger2006b}.  The statistic $X_{+}$ will have no B-mode 
contribution if $W_{-}(\ell)=0$ for all $\ell$ and thus would be an E-mode statistic. 
Similarly, $X_{-}$ would be a B-mode statistic.  Setting $W_{-}(\ell)=0$, I can integrate over $\ell$ against $\ell\,J_{4}(\ell\phi)$ to get 
\begin{eqnarray}
\lefteqn{\sum_{i}F_{-i}\int_{L_{i}}^{H_{i}}d\theta\,W_{i}(\theta)\int_{0}^{\infty} d\ell\,\ell\,J_{4}(\ell\theta)\,J_{4}(\ell\phi)}&&\nonumber\\
&&=\sum_{i}F_{+i}\int_{L_{i}}^{H_{i}}d\theta\,W_{i}(\theta)\int_{0}^{\infty} d\ell\,\ell\,J_{0}(\ell\theta)\,J_{4}(\ell\phi)\nonumber\ .
\end{eqnarray}
This equation must hold for arbitrary $\phi$.  Now let $\phi\in[L_{k},H_{k}]$.  Using the closure relationship for the Bessel functions 
\begin{equation}
\int_{0}^{\infty}d\ell\,\ell\,J_{4}(\ell\theta)\,J_{4}(\ell\phi) = \frac{1}{\phi}\delta(\theta-\phi)\nonumber
\end{equation}
and the following result from \citet{schneider2002}
\begin{eqnarray}
G(\theta,\phi) &=&\int_{0}^{\infty}d\ell\,\ell\,J_{0}(\ell\theta)\,J_{4}(\ell\phi)\nonumber\\
&=& \left(\frac{4}{\phi^{2}} - \frac{12\theta^{2}}{\phi^{4}}\right)H(\phi-\theta) + \frac{1}{\phi}\delta(\phi-\theta)\nonumber\ ,
\end{eqnarray}
I get the following expression for $F_{-k}$ in terms of the $F_{+i}$
\begin{eqnarray}
\lefteqn{F_{-k} = F_{+k}+\frac{\phi}{W_{k}(\phi)}}\nonumber\\
&&\ \ \ \times\sum_{i\leq k}F_{+i}\int_{L_{i}}^{{\rm min}(H_{i},\phi)}d\theta\,W_{i}(\theta)\left(\frac{4}{\phi^{2}} - \frac{12\theta^{2}}{\phi^{4}}\right)\,.\label{eqn:fmnobin}
\end{eqnarray}
Here $H(\phi-\theta)$ is the Heaviside step function and $\delta(\phi-\theta)$ is the Dirac delta function.  
This equation is the discrete analog of the results presented in \citet{schneider2007}.  In fact, it can be obtained by setting the quantities $T_{\pm}(\phi)$ from 
 \citet{schneider2007} to $W_{i}(\phi)F_{\pm i}/\phi$ in each bin $[L_{i},H_{i}]$.

\begin{figure*}
\begin{center}
\includegraphics{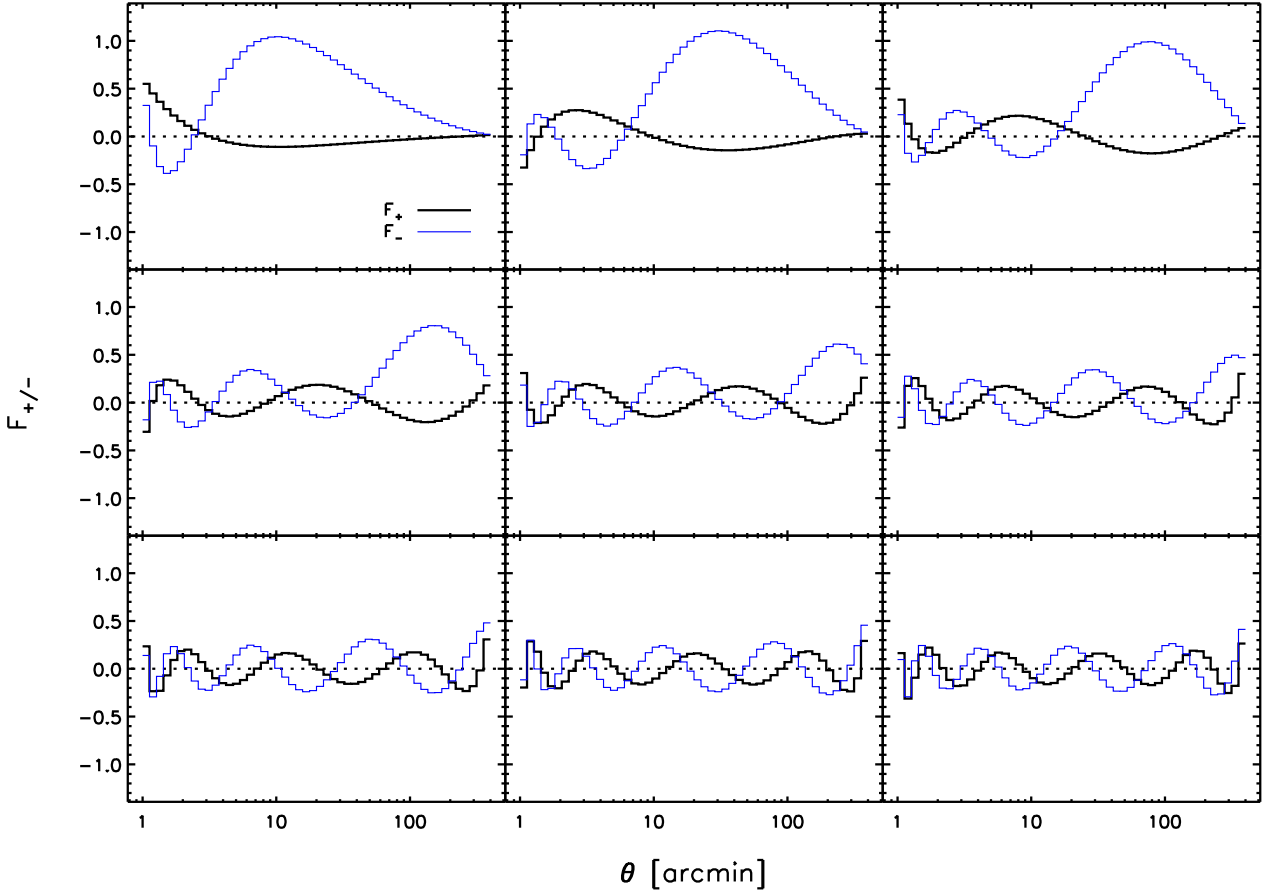}
\end{center}
\caption[]{The first nine components of the COSEBI-like basis for 50 shear correlation function 
data points with $\theta\in[1,400]$ arcmin. The thick black lines show the $F_{+}$ filters and the thin blue lines show the $F_{-}$ filters.  Each set of $F_{\pm}$ filters 
has been normalized to the same scale in each panel.\label{fig:cosebisbasis}}
\end{figure*}

\begin{figure*}
\begin{center}
\includegraphics[scale=0.48]{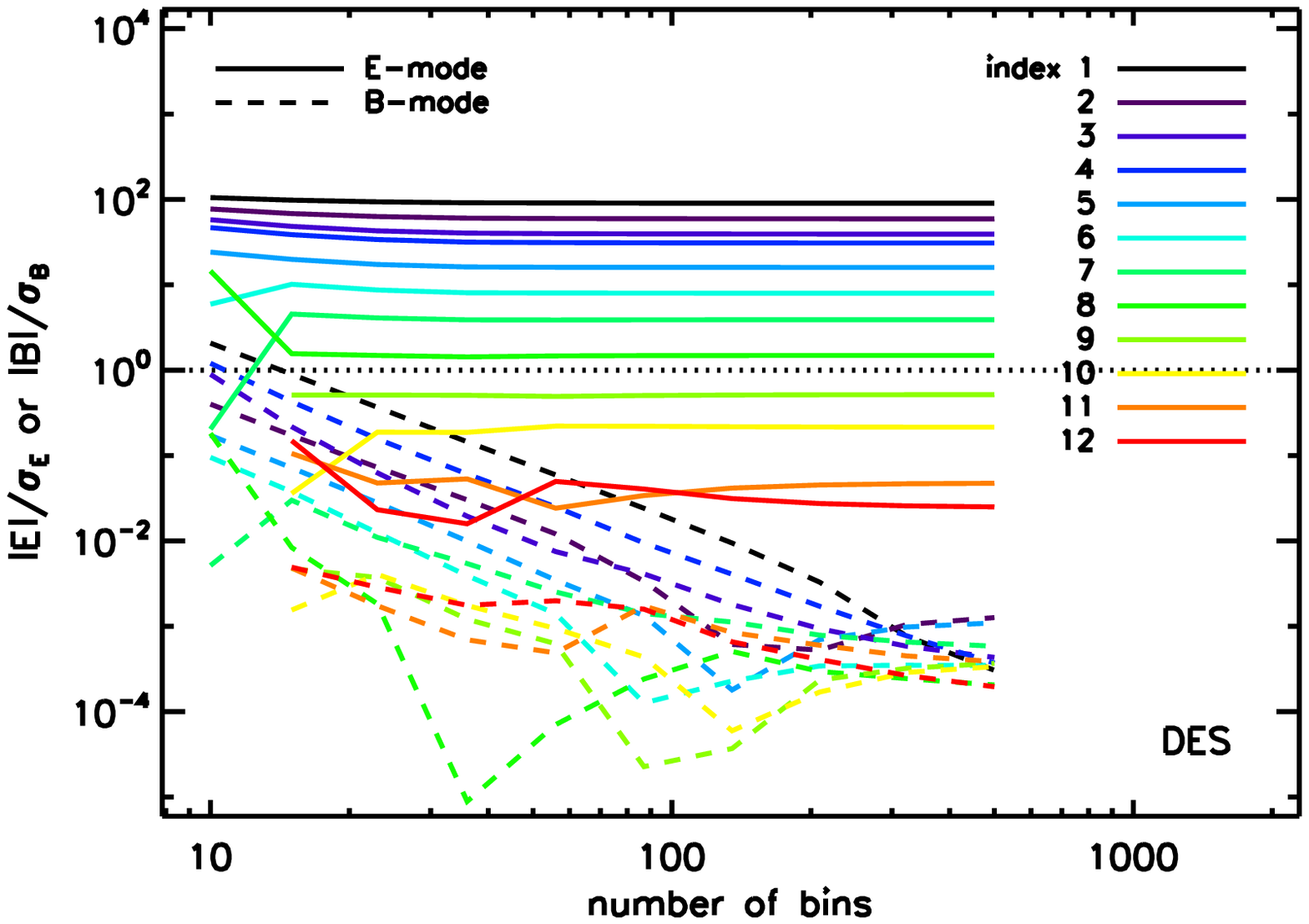}
\includegraphics[scale=0.48]{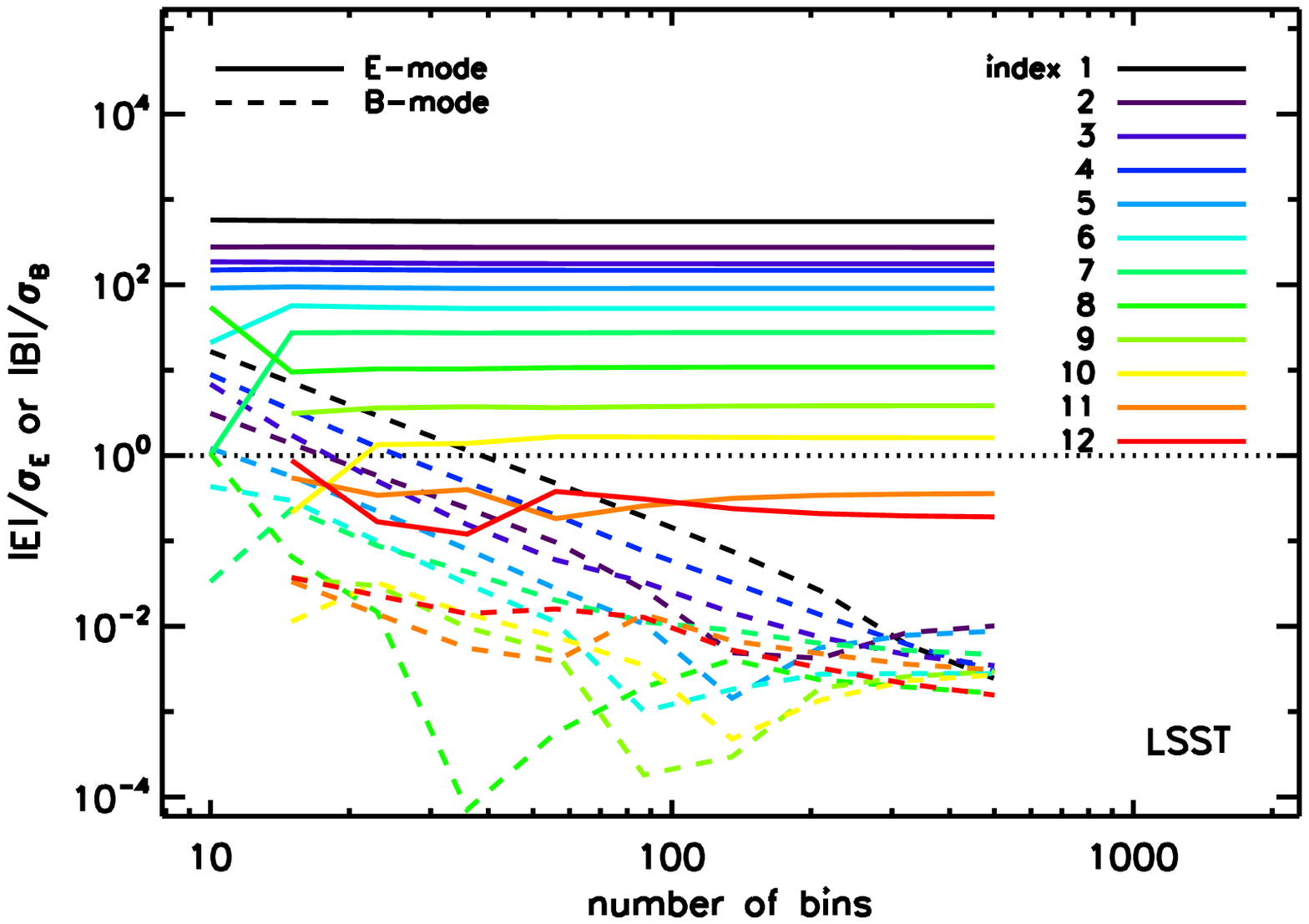}
\end{center}
\caption[]{The signal-to-noise of the COSEBI-like E- and B-mode statistics as a function of the number of shear correlation function bins.  The solid lines show from top to bottom 
the signal-to-noise of the first 12 E-mode statistics (estimated using only the diagonal elements of the covariance matrix).  The dashed lines show the signal-to-noise in the B-mode 
statistics.  The dotted line in each panel marks a signal-to-noise of unity.  The left panel is for a DES-like survey, while the right panel is for an LSST-like survey.  The intrinsic B-mode power 
was set to zero for this computation, so any statistically significant B-mode statistics are due purely to E/B-mode mixing.  This mixing decreases as the number of shear 
correlation function bins increases.  Additionally, the fact that only $\sim\!8-10$ of the E-mode statistics are statistically significant illustrates the data compression properties of these statistics.
\label{fig:cosebiss2n}}
\end{figure*}

Originally I assumed that the $F_{\pm i}$ were constants not functions of $\phi$, but the equation derived above in general has dependence on 
$\phi$.  Therefore generally $W_{-}(\ell)$ will not be identically zero for all $\ell$. The only way to ensure $W_{-}(\ell)\equiv 0$ is to pick the bin window 
functions $W_{i}(\phi)$ so that they exactly cancel the $\phi$ dependence in the equation above and then use this 
equation to compute the $F_{-i}$ in terms of the $F_{+i}$. Apart from this caveat, statistics of the form $X_{\pm}$ will mix E- and 
B-mode information in general. This mixing arises directly from the binning and is a general feature of E/B-mode separation with pixelized data as well \citep[see e.g.,][]{smith2006,lin2011}.

These results have important implications for the theoretical analysis of binned cosmic shear correlation function data.  For binned shear correlation functions with 
$N$ bins, it would be nice to divide the space of the 2$N$ data points into pure E-mode, pure B-mode, and ambiguous modes, similar to the decomposition 
achieved by \citet{schneider2010} with the COSEBI statistics over the continuous function space of the shear correlation functions.  However, I have shown that for general 
window functions $W_{i}(\theta)$, this decomposition is impossible because there do not exist pure E- and B-mode linear combinations.  In the next section, I will define 
approximately pure E- and B-mode linear combinations, along with approximately ambiguous modes.  Thus I will show that there does in fact exist a division of the space of 
$2N$ data points into \textit{approximately} ambiguous, E- and B-mode subspaces.  This approximate decomposition is explored fully in Section~\ref{sec:ebmodedec}.

These results are also quite useful for cosmic shear data analysis.  
Suppose one did in fact use a statistic which is a linear combination of the shear correlation function data points. Then from 
Equations (\ref{eqn:xipmell}) and (\ref{eqn:wpmdef}) one can compute how the statistic mixes E- and B-modes due to the binning. Additionally as I will show below, as the shear correlation 
function bins are made smaller, the magnitude of $W_{-}(\ell)$ will decrease, so that the E/B-mode mixing decreases as well.  If the bins are 
made small enough, the bias in $X_{-}$, the B-mode statistic, due to E/B-mode mixing, can be made smaller 
than the statistical errors.  Thus these results give a quantitative criterion by which to decide the number and size of the bins used to compute 
the shear correlation functions. Finally, Equation (\ref{eqn:fmnobin}) suggests two ways to minimize the E/B-mode mixing.  The first is to reweight the data in 
each bin by choosing the $W_{i}(\phi)$ to cancel the $\phi$ dependence in Equation~(\ref{eqn:fmnobin}) and then use this equation to compute the $F_{-i}$. 
I will not consider this possibility here. The second is to pick the $F_{-i}$ in order to minimize the E/B-mode mixing without adjusting the bin window functions.  
Given a fiducial choice for the $F_{+i}$, one can define $F_{-i}$ in some way (roughly similar to Equation~(\ref{eqn:fmnobin})) in order to minimize the mode mixing by minimizing 
$W_{-}(\ell)$.  The details of this definition along with a general procedure for computing the $F_{\pm i}$ are described in the next section.

\subsection{Building Binned E/B-mode Estimators}\label{sec:buildEBest}
Consider now binned shear correlation function data over the range $[L,H]$ in angle with $N$ bins.  Each bin is described by the bin window functions introduced above so that 
the bins cover the range $[L,H]$ without any overlaps between the ranges of each bin, $[L_{i},H_{i}]$.  In order to define the $F_{-k}$, I minimize the square amplitude of the window function $W_{-}(\ell)$ with respect to the coefficients $F_{-k}$,
\begin{equation}\label{eqn:optdef}
0=\frac{\partial}{\partial F_{-k}}\left[\int_{0}^{\infty}d\ell\,\ell|W_{-}(\ell)|^{2}\right]\ .
\end{equation}
The solution to this equation is
\begin{eqnarray}
\lefteqn{F_{-k}=F_{+k}  + \left(\int_{L_{k}}^{H_{k}}d\theta\frac{W_{k}^{2}(\theta)}{\theta}\right)^{-1}\sum_{i}F_{+i}}\nonumber\\
&&\times\int_{L_{i}}^{H_{i}}\int_{L_{k}}^{H_{k}}d\theta\,d\phi\,W_{i}(\theta)\,W_{k}(\phi)\nonumber\\
&&\ \ \ \ \  \ \ \ \ \ \ \ \ \ \ \ \ \ \ \ \ \ \ \ \ \ \ \ \ \ \ \ \times\left(\frac{4}{\phi^{2}} - \frac{12\theta^{2}}{\phi^{4}}\right)H(\phi-\theta)\ .\nonumber\\
&&\label{eqn:fmdef}
\end{eqnarray}
This solution is equivalent to multiplying Equation~(\ref{eqn:fmnobin}) by $W_{k}(\phi)/\phi$ and then integrating over $\phi$ with the weight $W_{k}(\phi)$.  

Similarly to the unbinned estimators described in \citet{schneider2007}, the binned estimators described in this work must satisfy two integral constraints in order to be non-zero 
only over a finite range in angle.  These constraints can be understood as follows.  Suppose that the $F_{+i}$ are non-zero only in the interval $[L_{s},H_{s}]$ contained in 
$[L,H]$.  Now consider $\phi\geq H_{s}$.  Then according to Equation~(\ref{eqn:fmdef}), the $F_{-i}$ will be non-zero in the interval $[L_{s},H_{s}]$ only if the $F_{+i}$ satisfy the 
following constraints
\begin{eqnarray}
0 &=& \sum_{\theta_{i}\in[L_{s},H_{s}]}F_{+i}\int_{L_{i}}^{H_{i}}d\theta\,W_{i}(\theta)\label{eqn:sumconta}\\
0 &=& \sum_{\theta_{i}\in[L_{s},H_{s}]}F_{+i}\int_{L_{i}}^{H_{i}}d\theta\,W_{i}(\theta)\,\theta^{2}\label{eqn:sumcontb}
\end{eqnarray}
where the sums run only over bins in the interval $[L_{s},H_{s}]$.  Given data in some fiducial angular interval, one can always increase the angular range considered 
as long as the $F_{\pm i}$ are zero outside the fiducial angular range.  Thus in order for the $X_{\pm}$ statistics to self-consistently consider only data in 
a finite angular range and minimize E/B-mode mixing by minimizing $W_{-}(\ell)$, they must satisfy the constraints given above. 

As discussed in \citet{schneider2010}, these constraints serve to project out ambiguous modes in the continuous correlation functions $\xi_{\pm}$ which cannot be uniquely classified as 
either E- or B-modes.  These modes are $\xi_{+}(\theta)=a+b\theta^{2}$ and so these constraints simply project out the binned versions of these modes.  Barring the caveat discussed above, 
in the discrete case no modes can be uniquely classified as pure E- or B-modes.  Thus in this sense all modes are ambiguous in the discrete case, due to the binning.  However, 
E/B-mode separation for the discrete modes along $\xi_{+}(\theta)=a+b\theta^{2}$ is not possible even in the limit of infinitely small bins, so these discrete modes are the 
analogues of the ambiguous continuous modes.  Importantly, these ambiguous modes can still potentially carry cosmological information, it is just that one cannot 
uniquely determine if the modes are sourced by the E- or B-mode power spectrum. Assuming one has determined that the cosmic shear data is free of systematic 
contamination, it may be advantageous to keep and use the information in the ambiguous modes (see \citealt{smith2006} for a similar point in the context of the 
pseudo-$C_{\ell}$ formalism and also \citealt{schneider2010}).  This point is illustrated below with a simple Fisher information analysis.

Given the linear relation between the $F_{\pm i}$ and the linear constraints on the $F_{+i}$, it is natural to treat them as vectors of length $N$.  
Thus the two integral constraints in Equations~(\ref{eqn:sumconta}) and (\ref{eqn:sumcontb}) mean that the $F_{+}$ vector cannot have any component along the directions 
\begin{eqnarray}
F_{+a} &=& \left(\int_{L_{1}}^{H_{1}}d\theta\,W_{1}(\theta),\int_{L_{2}}^{H_{2}}d\theta\,W_{2}(\theta),...,\right.\nonumber\\
&&\left.\ \ \ \ \int_{L_{N}}^{H_{N}}d\theta\,W_{N}(\theta)\right)_{N}\nonumber\\
F_{+b} &=& \left(\int_{L_{1}}^{H_{1}}d\theta\,W_{1}(\theta)\,\theta^{2},\int_{L_{2}}^{H_{2}}d\theta\,W_{2}(\theta)\,\theta^{2},...,\right.\nonumber\\
&&\left.\ \ \ \ \int_{L_{N}}^{H_{N}}d\theta\,W_{N}(\theta)\,\theta^{2}\right)_{N}\nonumber\ .
\end{eqnarray}
Similarly, Equation~(\ref{eqn:fmdef}) defines a matrix, which I will denote as $\mathrm{M}_{+}$, which relates the $F_{+}$ to the $F_{-}$ via $F_{-}=\mathrm{M}_{+}F_{+}$.  I give explicit forms 
for these quantities assuming geometric bin weightings in Appendix~\ref{app:estgeomwgt}.  

The overall process to create a set of $F_{+}$ and $F_{-}$ vectors and measure the E- and B-mode signals in this scheme is quite straight forward.
\begin{enumerate}
\renewcommand{\theenumi}{(\arabic{enumi})}
\item Choose an initial shape for the $F_{+}$ vector.  Usually one has some external motivation for this choice, like for example, statistics which form a complete set of functions 
and exhibit efficient data compression \citep{schneider2010}, statistics that are compact and non-oscillatory in Fourier space \citep{leonard2012}, or 
statistics which maximize the signal-to-noise in the measurement of cosmological parameters, like $\sigma_{8}$ or $\Omega_{m}$ \citep{fu2010}.
\item Make $F_{+}$ is orthogonal to the $F_{+a}$ and $F_{+b}$ vectors defined above.  In practice, an efficient way to do this is to first make
$F_{+a}$ and $F_{+b}$ orthogonal to one another by modifying $F_{+b}$.  Then it is easy to make $F_{+}$ orthogonal to both $F_{+a}$ and the modified
$F_{+b}$ by projecting out the components of $F_{+}$ along the modified constraint directions.
\item Use the matrix $\mathrm{M}_{+}$ defined in Equation (\ref{eqn:fmdef}) to compute the corresponding $F_{-}$ filter via $F_{-}=\mathrm{M}_{+}F_{+}$.
\item Compute the E- and B-mode signals by forming the vector inner product of the binned $\widehat{\xi}_{\pm}$ shear correlation function
data with the $F_{\pm}$ filters.  Then $E=X_{+}=(I_{+}+I_{-})/2$ and $B=X_{-}=(I_{+}-I_{-})/2$, where
$I_{\pm}=\sum_{i}\widehat{\xi}_{\pm i}F_{\pm i}$.
\end{enumerate}

Below I use both the procedure just described and also Gram-Schmidt orthogonalization to define $F_{+}$ functions.  In the case of Gram-Schmidt
orthogonalization, one simply uses $F_{+a}$ and $F_{+b}$ first in the Gram-Schmidt process, followed by the rest of the initial estimator shapes for $F_{+}$.  The
result of applying the Gram-Schmidt procedure is a set of orthogonal $F_{+}$ filters contained in the $N-2$ elements of the orthogonal basis.  The
first two elements of this basis span the space generated by the constraint directions $F_{+a}$ and $F_{+b}$.  The $F_{-}$ filters are again obtained
from the matrix defined in Equation (\ref{eqn:fmdef}).  

\subsection{Other Potential Optimal Estimator Definitions}\label{sec:otherest}
In this section I will explore two other potential optimal estimators definitions.  The first definition simply consists of swapping the roles of $F_{+}$ and $F_{-}$ in the 
previous section.  Thus in this scheme the optimal estimator is now defined by 
\begin{equation}
0 =\frac{\partial}{\partial F_{+k}}\left[\int_{0}^{\infty}d\ell\,\ell|W_{-}(\ell)|^{2}\right]\ .
\end{equation}
The full estimators for this case are presented in Appendix~\ref{app:fminusest}.  In this case, all of the same results presented above hold and in particular there 
exists a matrix $\mathrm{M}_{-}$ which defines the $F_{+}$ in terms of the $F_{-}$ via $F_{+}=\mathrm{M}_{-}F_{-}$.  There also exist ambiguous modes, $F_{-a}$ and $F_{-b}$, in analogy to the ones 
defined above.  Finally, the procedure just given for E/B-mode separation can be carried out in exactly the same way as described above by first defining fiducial $F_{-}$ filters, 
projecting out the $F_{-a}$ and $F_{-b}$ modes, computing the $F_{+}$ filters from the matrix $\mathrm{M}_{-}$, and then computing $X_{\pm}$.

The second potential optimal estimator definition consists of minimizing the power in the $W_{-}(\ell)$ window function by varying both $F_{+}$ and $F_{-}$ simultaneously.  
The two equations that must be solved for this kind of estimator are 
\begin{eqnarray}
0 &=&\frac{\partial}{\partial F_{-k}}\left[\int_{0}^{\infty}d\ell\,\ell|W_{-}(\ell)|^{2}\right]\nonumber\\
0 &=&\frac{\partial}{\partial F_{+k}}\left[\int_{0}^{\infty}d\ell\,\ell|W_{-}(\ell)|^{2}\right]\nonumber\ .
\end{eqnarray}
Using the results for the two optimal estimators already presented, the two equations the $F_{\pm}$ vectors must satisfy are
\begin{eqnarray}
F_{-} &=& \mathrm{M}_{+}\,F_{+}\nonumber\\
F_{+} &=& \mathrm{M}_{-}\,F_{-}\nonumber\ .
\end{eqnarray}
Combing these two relations one gets that the optimal $F_{+}$ and $F_{-}$ vectors computed in this way must satisfy
\begin{eqnarray}
0 &=& (\mathrm{M}_{-}\mathrm{M}_{+} - \mathrm{I})\,F_{+}\nonumber\\
0 &=& (\mathrm{M}_{+}\mathrm{M}_{-} - \mathrm{I})\,F_{-}\nonumber\ ,
\end{eqnarray}
where $\mathrm{I}$ is the identity matrix.  Thus in this case the $F_{\pm}$ vectors must be in the null space or kernels of 
the matrices $\mathrm{M}_{\mp}\mathrm{M}_{\pm} - \mathrm{I}$.

\begin{figure*}
\begin{center}
\includegraphics{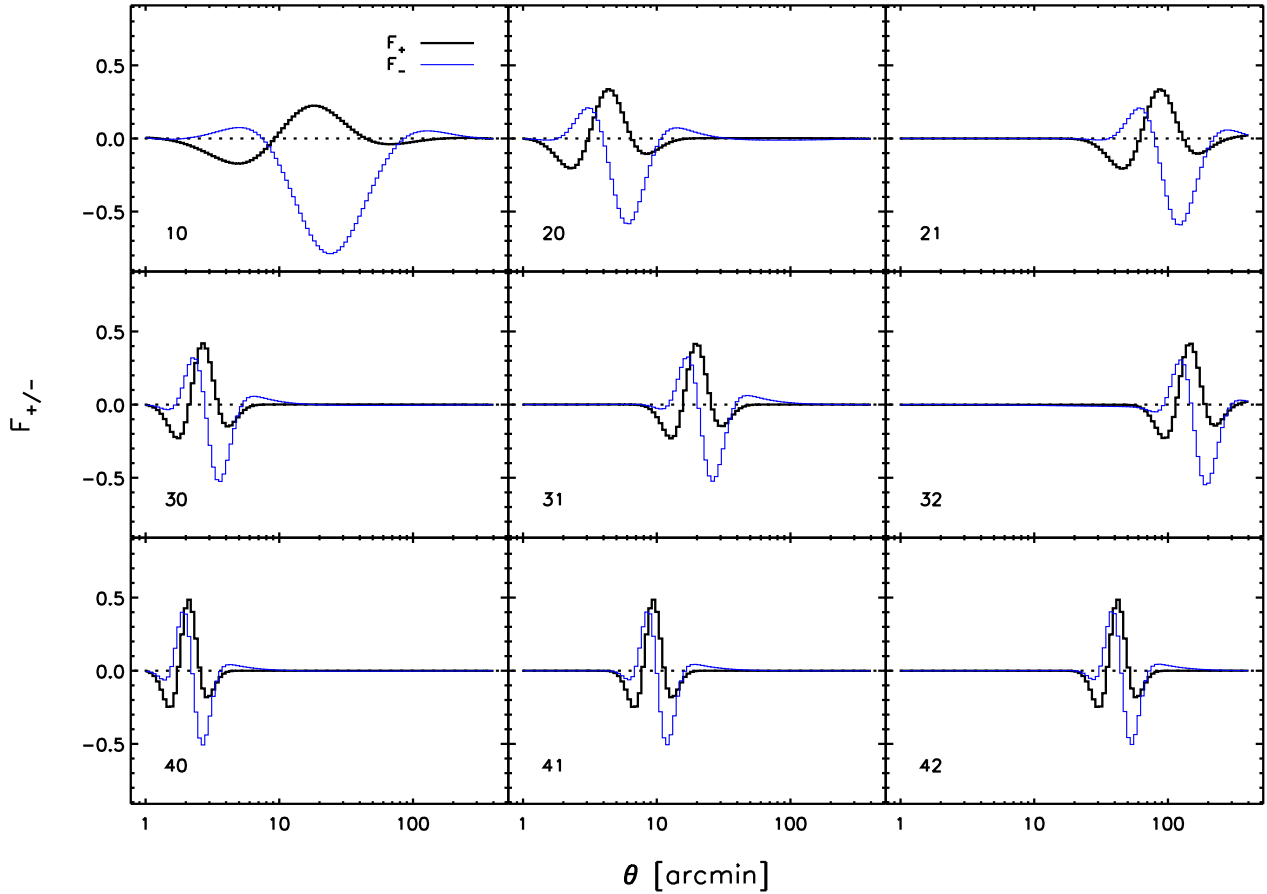}
\end{center}
\caption[]{The first nine components of the wavelet-like estimators for 100 correlation function 
data points and $\theta\in[1,400]$ arcmin. The thick black lines show the $F_{+}$ filters and the blue lines show the $F_{-}$ filters.  
The index $\{N,i\}$ of each filter is given in the lower left corner of each panel. \label{fig:hatwavebasis}}
\end{figure*}

A straight forward way to investigate the kernels of these matrices is via computing their Singular Value Decomposition (SVD) \citep[e.g.,][]{press1992}.  The SVD of a matrix $\mathrm{M}$ is defined 
as $\mathrm{M}=\mathrm{U}\Sigma\mathrm{V}^{T}$. Here $\mathrm{U}$ and $\mathrm{V}$ are orthogonal matrices, assuming $\mathrm{M}$ is real, and $\Sigma$ is a 
diagonal matrix with the singular values along the diagonal.  One can show that a basis for the kernel of a square matrix can be found by using all of the columns of $\mathrm{V}$ 
which have zero singular values in $\Sigma$.  Similar results hold for non-square matrices but they are of no use here.  Additionally, the rank of the matrix is number of non-zero 
singular values so that a square matrix with all non-zero singular values has full rank and is invertible.

Using logarithmic binning and the geometric bin weighting $W_{i}(\theta)$, I find that the matrix $\mathrm{M}_{-}\mathrm{M}_{+} - \mathrm{I}$ has two large singular values which exceed all of the others 
by several orders of magnitude. The $N-2$ other singular values are of similar magnitude to each other, but much smaller than the largest two.  The difference between the two largest and the other $N-2$ 
singular values increases as the number of bins is increased in a fixed angular range.  Also, the columns of $\mathrm{V}$ along the two largest singular values are approximately along the 
ambiguous directions $F_{+a}$ and $F_{+b}$.  Analogous properties hold for the singular values of the matrix $\mathrm{M}_{+}\mathrm{M}_{-} - \mathrm{I}$ and the ambiguous directions 
$F_{-a}$ and $F_{-b}$.  Unfortunately, all of the singular values of these matrices are non-zero so that no vector except the zero vector lies in their null spaces.  
Note that one must be careful to account for rounding errors when considering whether or not singular values are zero.  I find however that the $N-2$ smaller singular values 
are well above rounding errors.  Thus optimal estimators defined by varying both the $F_{+}$ and $F_{-}$ coefficients simultaneously do not exist in this case.

However, this analysis does provide useful insight into the relationship between the estimators defined by varying $F_{-}$ and those defined by varying the $F_{+}$ coefficients.  In particular, given 
that the singular values along the directions orthogonal to the ambiguous directions $\{F_{+a},F_{+b}\}$ and $\{F_{-a},F_{-b}\}$ are very small, we have that $F_{+}$ and $F_{-}$ vectors orthogonal to 
these directions span subspaces which are close to being kernels of the matrices $\mathrm{M}_{\mp}\mathrm{M}_{\pm} - \mathrm{I}$.  Thus 
\begin{eqnarray}
0 &\approx& (\mathrm{M}_{-}\mathrm{M}_{+} - \mathrm{I})\,F_{+}\nonumber\\
0 &\approx& (\mathrm{M}_{+}\mathrm{M}_{-} - \mathrm{I})\,F_{-}\nonumber\ .
\end{eqnarray}
One can verify this property empirically as well.  Therefore the matrix $\mathrm{M}_{+}$ generates $F_{-}$ vectors which are 
approximately in the approximate null space of $\mathrm{M}_{+}\mathrm{M}_{-} - \mathrm{I}$ and vice versa.  (To see this, simply multiply the two equations above by $\mathrm{M}_{+}$ 
and $\mathrm{M}_{-}$ respectively and use the relationship $F_{\mp} = \mathrm{M}_{\pm}F_{\pm}$.) Thus $F_{-}$ vectors generated from $F_{+}$ vectors are very close to being orthogonal 
to the constraint directions $\{F_{-a},F_{-b}\}$.  The analogous property holds for $F_{+}$ vectors generated from $F_{-}$ vectors and the constraint directions $\{F_{+a},F_{+b}\}$.  
Additionally, due to these approximate equalities, the operation of ``composing'' the two different estimator definitions approximately returns the identity.  
In other words, one can first compute optimal estimators by varying the $F_{-}$ and supplying fiducial guesses for the $F_{+}$.  Then the computed $F_{-}$ vectors can be used to supply 
fiducial guesses for the set of optimal estimators defined by varying the $F_{+}$.  The new set of estimators that result from varying the $F_{+}$ will be very close to the first set of 
estimators defined by varying the $F_{-}$.  This fact can be verified empirically.  In this sense, the two optimal estimator definitions are consistent.

\subsection{Approximate E-, B-, and Ambiguous Mode Decompositions for Binned Cosmic Shear Data}\label{sec:ebmodedec}
It is now straight forward to build the approximate decomposition of the space of $2N$ data points into E-mode, B-mode, and 
ambiguous modes discussed above. To do this properly, I must specify a basis for the vector space of $2N$ data points which can be divided into directions approximately 
along ambiguous, E- and B-modes. The subspaces spanned by each of these three sets of modes need not be mutually orthogonal, though in practice the E- and B-mode 
subspaces are approximately orthogonal to the ambiguous mode subspace defined below.

\begin{figure*}
\begin{center}
\includegraphics[scale=0.48]{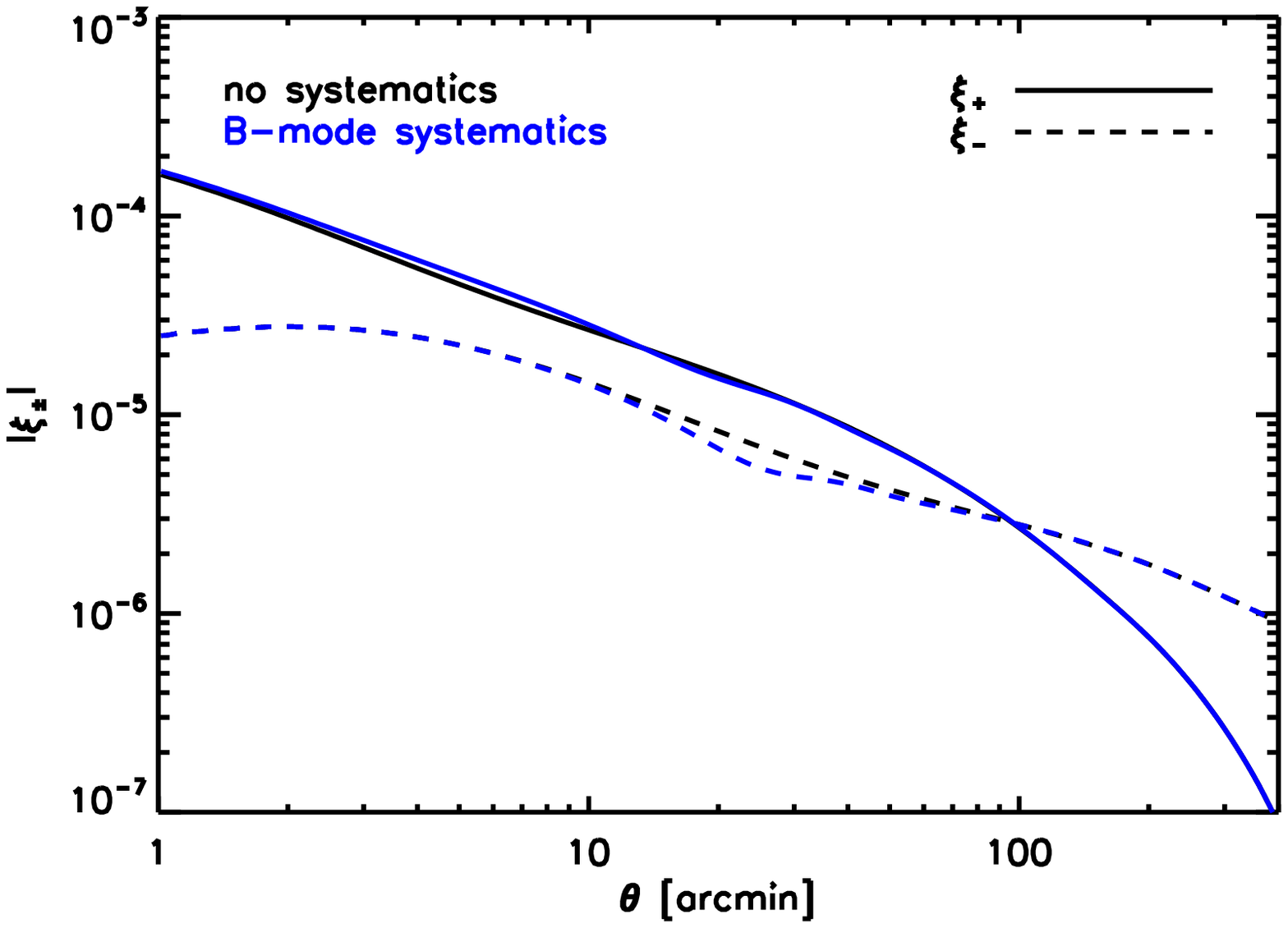}
\includegraphics[scale=0.48]{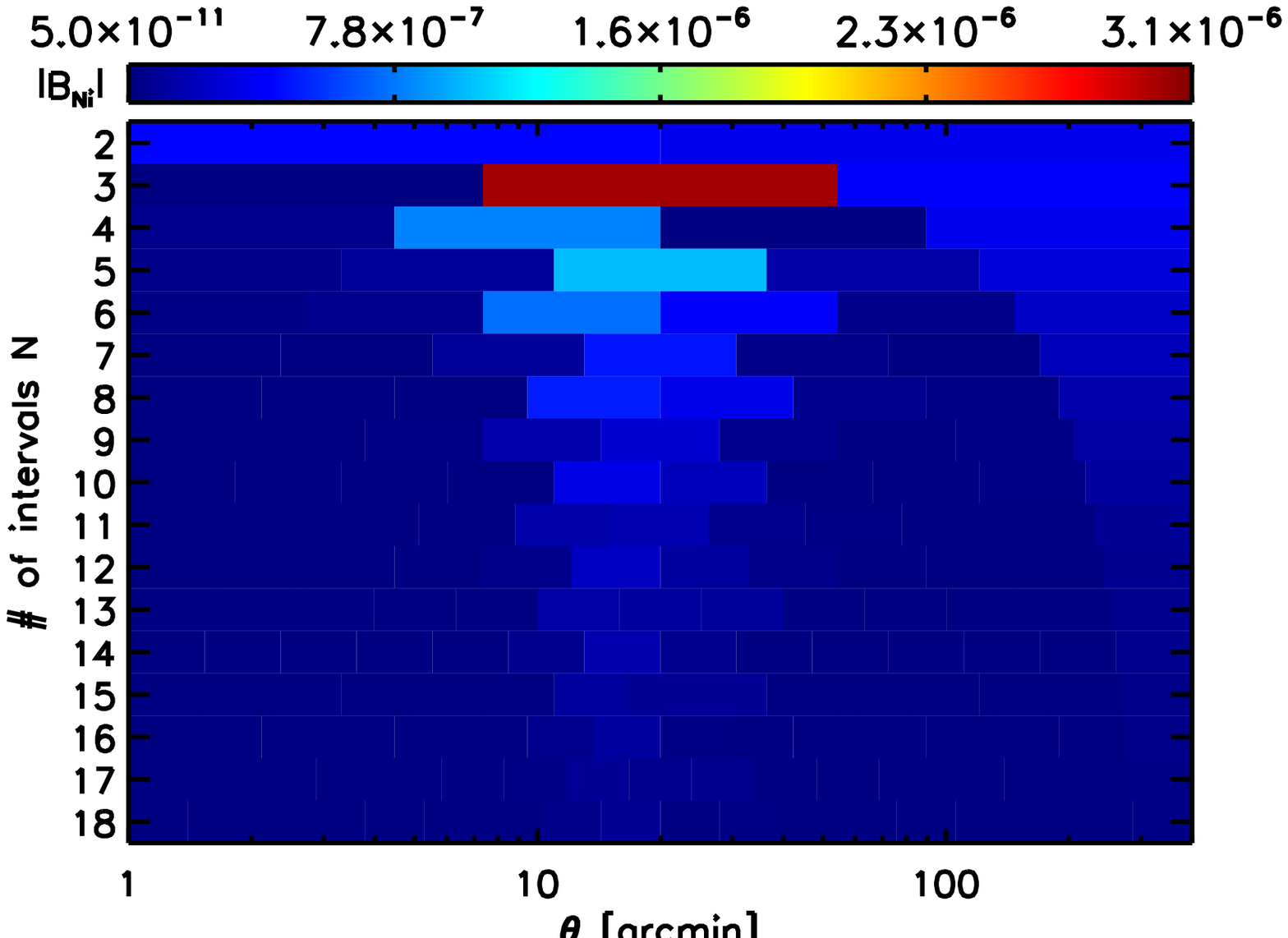}
\includegraphics[scale=0.48]{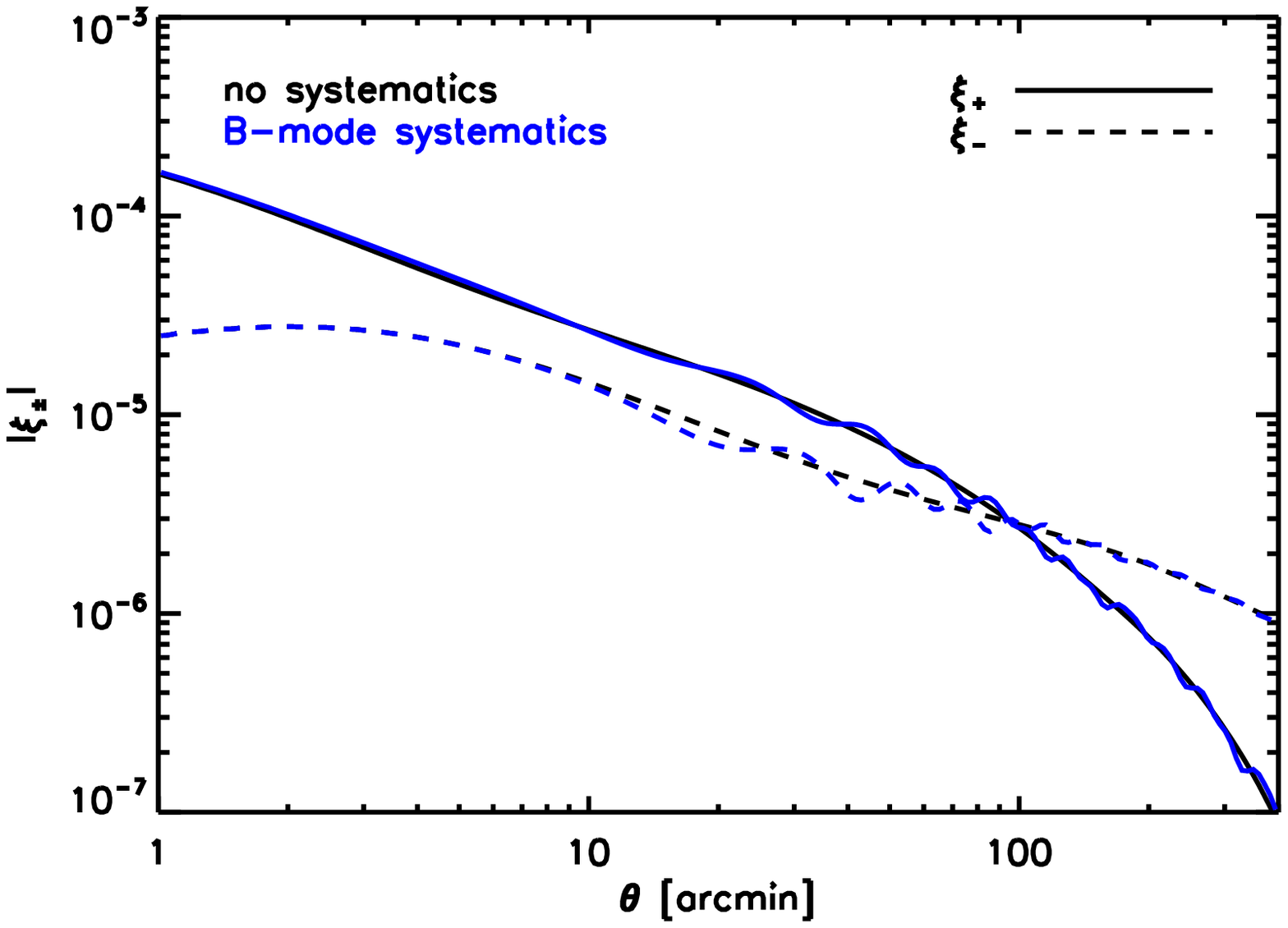}
\includegraphics[scale=0.48]{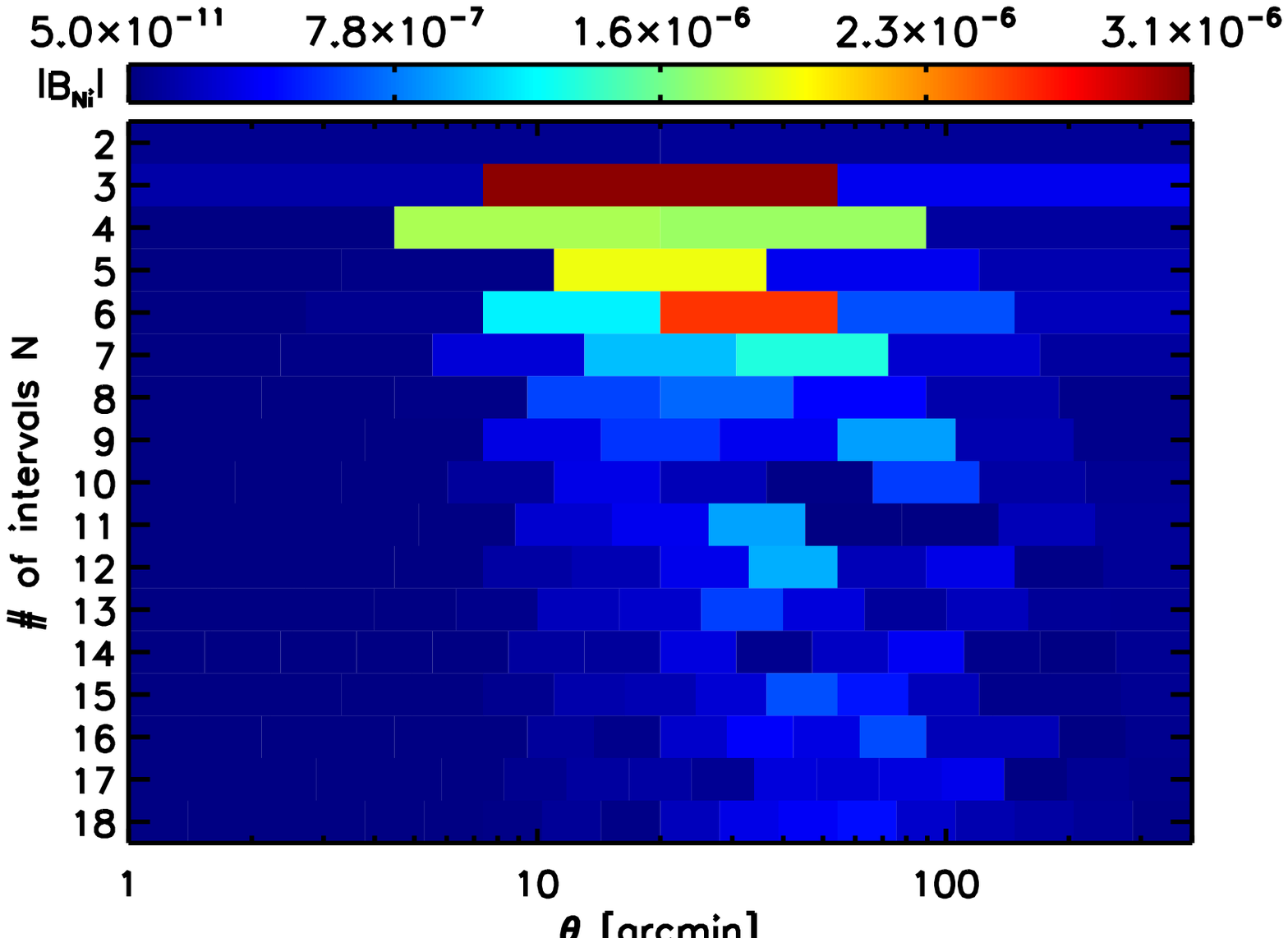}
\end{center}
\caption[]{Example B-mode detection in noise-free data with 200 correlation function points from 1 to 400 arcmin.  Each row on the left shows the
underlying true correlation functions in black and the correlation functions with systematics in blue.  On the right each row shows the
absolute value of the wavelet-like scale-location B-mode statistics defined in Section \ref{sec:wvstats}.  The vertical axis label denotes how many
sections the interval $[1,400]$ was divided into in log-space with the intervals increasing in size from bottom to top.  The interval width is
reflected in the size of each colored region.  The color of each region encodes the amplitude of the B-mode level.  The top row has a smooth
systematic signal while the bottom row has much more smaller scale variations.  These finer scale
variations result in B-mode detections for the smaller scale statistics for the bottom row, while no such detections are found for the top
row.\label{fig:wvtest}}
\end{figure*}

First, I specify the basis vectors for the ambiguous directions.  These vectors are
\begin{equation}\nonumber
\begin{array}{l}
\{F_{+a},\vec{0}\}\\
\{F_{+b},\vec{0}\}\\
\{\vec{0},F_{-a}\}\\
\{\vec{0},F_{-b}\}
\end{array}
\end{equation}
where the $\vec{0}$ are zero vectors of length $N$. Next I specify the E- and B-mode directions.  This is done by building a set of $N-2$ E- and B-mode estimators from 
a set of $N-2$ $F_{+}$ vectors.  Additionally these $N-2$ $F_{+}$ vectors together with $F_{+a}$ and $F_{+b}$ should form a basis for the $\xi_{+}$ subspace of the total 
vector space composed of both $\xi_{+}$ and $\xi_{-}$.  The construction of the optimal estimators enforces that the $N-2$ $F_{+}$ vectors are exactly orthogonal to the 
$F_{+a}$ and $F_{+b}$ vectors.  Additionally, as shown above, the resulting $F_{-}$ vectors will be approximately orthogonal 
to the $F_{-a}$ and $F_{-b}$ vectors.  So using the $N-2$ $F_{+}$ vectors along with their $F_{-}$ counterparts, one can define the 
final $2N-4$ potential basis vectors of this space as 
\begin{equation}\nonumber
\begin{array}{l}
\{F_{+},+F_{-}\}/2\\
\{F_{+},-F_{-}\}/2\ ,
\end{array}
\end{equation}
where this construction is repeated for each of the $N-2$ $F_{+}$ vectors.  The first of these vectors is simply the direction of $X_{+}$ and the second is in the direction of 
$X_{-}$ so that they are approximate E- and B-mode directions.  Finally, one must verify that this set of vectors combined with the ambiguous mode directions is in fact a basis for 
the vectors space of dimension $2N$ by, for example, determining the rank of the matrix composed of these vectors as rows.  Note that this construction would 
work just as well by first building a basis with the $F_{-}$ vectors and the $\{F_{-a},F_{-b}\}$ vectors, and then computing the $F_{+}$ vectors.  Below, I will investigate 
this construction for the COSEBI-like \citep{schneider2010} basis vectors defined in Section~\ref{sec:cosebi}.

Finally, it is important to note that the exact properties of the matrices $\mathrm{M}_{\pm}$ influence the properties of the basis constructed above nontrivially.  Specifically, if one uses 
the equal area binning (i.e. $H_{i}^{2}-L_{i}^{2}$ is the same for all bins $i$), then one can show using the expressions for these matrices given in Appendices~\ref{app:estgeomwgt} 
and \ref{app:fminusest} for geometric binning functions that $\mathrm{M}_{-}=\mathrm{M}_{+}^{T}$.  Thus for vectors along the $F_{\pm}$ directions, the matrices $\mathrm{M}_{\pm}$ in this 
case are approximately orthogonal.  A quick calculation shows that if these matrices were exactly orthogonal and one started out with an orthonormal basis for the $\xi_{+}$ subspace, then 
then the $F_{-}$ vectors would also be orthonormal.  In this case the E- and B-mode subspaces would then be exactly mutually orthogonal.  In practice this exact orthogonality is not realized, but the 
E- and B-mode subspaces do retain some degree of orthogonality even for logarithmic binning for the COSEBI-like statistics described below in Section~\ref{sec:cosebi}.

\subsection{The Binned E/B-mode Estimator Covariances}\label{sec:ebcov}
With Equation~(\ref{eqn:xipmell}) it is possible to compute the covariance of these statistics.  I follow the same method outlined in \citet{schneider2010} and get, assuming that the EB cross-power $P_{EB}(\ell)$ is zero and the power spectra are Gaussian,
\begin{eqnarray}
\lefteqn{\mathrm{Cov}(E_{n},E_{m})=}&&\nonumber\\
&&\frac{1}{\pi\Omega_{s}}\int_{0}^{\infty}d\ell\,\ell\left[W_{+}^{n}(\ell)W_{+}^{m}(\ell)\left(P_{E}(\ell)+\frac{\sigma_{e}^{2}}{\bar{n}}\right)^{2}\right.\nonumber\\
&&\ \ \ \ \ \ \ \ \ \ \ \ \ \ \ \ \ \ +\left.W_{-}^{n}(\ell)W_{-}^{m}(\ell)\left(P_{B}(\ell)+\frac{\sigma_{e}^{2}}{\bar{n}}\right)^{2}\right]
\end{eqnarray}
\begin{eqnarray}
\lefteqn{\mathrm{Cov}(B_{n},B_{m})=}&&\nonumber\\
&&\frac{1}{\pi\Omega_{s}}\int_{0}^{\infty}d\ell\,\ell\left[W_{+}^{n}(\ell)W_{+}^{m}(\ell)\left(P_{B}(\ell)+\frac{\sigma_{e}^{2}}{\bar{n}}\right)^{2}\right.\nonumber\\
&&\ \ \ \ \ \ \ \ \ \ \ \ \ \ \ \ \ \ +\left.W_{-}^{n}(\ell)W_{-}^{m}(\ell)\left(P_{E}(\ell)+\frac{\sigma_{e}^{2}}{\bar{n}}\right)^{2}\right]
\end{eqnarray}
\begin{eqnarray}
\lefteqn{\mathrm{Cov}(E_{n},B_{m})=}&&\nonumber\\
&&\frac{1}{\pi\Omega_{s}}\int_{0}^{\infty}d\ell\,\ell\left[W_{+}^{n}(\ell)W_{-}^{m}(\ell)\left(P_{E}(\ell)+\frac{\sigma_{e}^{2}}{\bar{n}}\right)^{2}\right.\nonumber\\
&&\ \ \ \ \ \ \ \ \ \ \ \ \ \ \ \ \ \ +\left.W_{-}^{n}(\ell)W_{+}^{m}(\ell)\left(P_{B}(\ell)+\frac{\sigma_{e}^{2}}{\bar{n}}\right)^{2}\right]\ .
\end{eqnarray}
These expressions reduce to those in \citet{schneider2010} when $W_{-}(\ell)\equiv0$.  Also note that the squared amplitude of the $W_{-}(\ell)$ window function 
controls the amount of excess variance in the E- and B-mode statistics due to E/B-mode mixing.  The procedure for defining $F_{-}$ introduced above serves to minimize 
this excess variance.  Similarly, by minimizing $W_{-}(\ell)$, the covariance between the E- and B-mode statistics is minimized.

\section{Examples and Tests of the Estimators}\label{sec:extest}
In this section, I will illustrate the ideas discussed above by constructing and evaluating the performance of two different E/B-mode statistics.  I first present
COSEBI-like \citep*[hereafter SEK10]{schneider2010} statistics in Section~\ref{sec:cosebi}.  These statistics have $\sim\!10\times$ data compression ratios for upcoming surveys 
which can greatly simply data analysis.  Using these estimators I will also evaluate the level of E/B-mode mixing as function of the number of shear correlation function bins.  
Then I present wavelet-like B-mode statistics which allow one to determine the scale and location of B-mode contamination in shear correlation function data in Section~\ref{sec:wvstats}. 
Finally in Section~\ref{sec:fishinfo}, I explore the properties of the approximate ambiguous, E- and B-mode decomposition discussed above and the Fisher information content of these 
subspaces using the COSEB-like statistics.

\subsection{Binned COSEBI-like Estimators}\label{sec:cosebi}
As demonstrated in SEK10, filter functions which are polynomials in $\ln\theta$ have very optimal data compression ratios.  Additionally, SEK10 built
a set of filter functions which are orthonormal over a finite interval (using a slightly different definition of orthonormality than the vector inner product used above). In order to generate
a similar set of filters for correlation functions measured with $N$ bins, I take as an initial shape for the $F_{+}$ filters,
$P_{\ell}(t)/\exp[(\ln\theta-\ln L)/2]$, where $P_{\ell}(t)$ is the Legendre polynomial of order $\ell$ and $t$ is simply the logarithm of the angle $\theta$ remapped to the appropriate interval, 
$t\equiv2(\ln\theta - (\ln H + \ln L)/2)/(\ln H - \ln L)$.\footnote{Note that one can discretize the SEK10 statistics directly using the correspondence 
between the $T_{\pm}$ and $F_{\pm}$ filter functions, $F_{\pm i}\sim T_{\pm}(\theta_{i})\theta_{i}/W_{i}(\theta_{i})$ where $\theta_{i}$ is a representative point in bin $i$.  
However, I have found that this procedure produces statistics which do not exhibit the data compression properties of the SEK10 statistics.} I increase $\ell$ for each vector in the basis
starting at $\ell=2$ and I use the logarithmic mean of each bin interval to evaluate the polynomial.  The first two vectors in the basis are the constraint directions $F_{+a}$ and $F_{+b}$ so that the 
third vector is started with $\ell=2$.  As the number of roots in the Legendre polynomials increases, they become harder to 
represent properly in the discrete binned space.  Thus when the index of the Legendre polynomial $\ell$ is greater than $2N/3$, where $N$ is the number of bins in the interval, I switch to generating 
the fiducial basis by producing $F_{+}$ vectors at order $\ell$ divided into $\ell+1$ sections alternating between $-1$ and $+1$.  This construction generates as many roots in the interval 
as the Legendre polynomial of order $\ell$ would have, but these roots are now resolved properly.  Note that the maximum value of $\ell$ is $N-1$, so that one gets exactly $N-1$ roots 
when $N$ intervals alternate between $-1$ and $1$.  Then the Gram-Schmidt procedure is applied to the constraint vectors and the $N-2$ fiducial $F_{+}$ vectors 
as described above in order to produce the final $F_{+}$ filters.  The factor of 
$\exp[(t-\ln L)/2]$ has been inserted to roughly account for the difference between the vector inner product definition and that in SEK10.  In Figure~\ref{fig:cosebisbasis}, I show the 
COSEBI-like vector basis for $F_{+n}$ and $F_{-n}$ for 50 shear correlation function data points between 1 and 400 arcmin.

In Figure \ref{fig:cosebiss2n}, I show the level of E/B-mode mixing for these statistics as a function of the number of shear correlation function bins.  I have plotted the ratio 
of the first 12 E- and B-mode statistics to their errors for both DES- and LSST-like surveys.  Note that in general these statistics are correlated, but to make this plot I only use 
the diagonal contributions to the error covariance matrix.  The statistics in this plot were computed with $P_{B}(\ell)\equiv0$, so any nonzero B-mode statistic amplitude is completely due 
to E/B-mode mixing.  Additionally, the E/B-mode mixing in the estimators is negligible if the nonzero amplitude of the B-mode statistic is well below the error bar on the 
statistic.  This condition ensures that any statistically significant systematic effect is detectable and is not confused with E-modes due to E/B-mode mixing.  

In general, the level of E/B-mode mixing decreases as the number of bins used for the shear correlation function increases.  Additionally, it is clear that for a DES-like survey 
fewer shear correlation function bins can be used than for an LSST-like survey.  If one uses a threshold of the B-mode statistic being no more than 10\% of the 1-sigma error bar, 
then a DES-like survey will need of order $\sim\!50$ bins whereas an LSST-like survey will need $\sim\!100$ bins.  
This figure also demonstrates the data compression properties of these statistics.  In the case of a DES-like survey of order 50 shear correlation function data points can be compressed 
into only $\sim\!8$ E-mode statistics which are measurably non-zero.  An LSST-like survey has increased statistical power, compressing of order 100 data points into only 
$\sim\!10$ statistically significant E-mode statistics.  Due to correlations between these remaining E-mode statistics, further data compression may be possible, but I will 
not explore this issue further in this work.  

\subsection{Wavelet-like B-mode Size-location Estimators}\label{sec:wvstats}
In this section, I give an example of the use of spatially-local basis functions to build sets of B-mode estimators which can pinpoint the size and
location of B-mode systematics in shear correlation function data.  Consider the following function, know as the Ricker wavelet \citep{ricker1953},
\begin{equation}
\psi(t,\sigma)=\frac{2}{\sqrt{3\sigma}\pi^{1/4}}\left(1-\frac{t^{2}}{\sigma^{2}}\right)\exp\left[-\frac{t^{2}}{2\sigma^{2}}\right]\ .
\end{equation}
I take as a set of starting functions the following functions, indexed by $\{i,N\}$,
\begin{eqnarray}
\lefteqn{\phi_{iN}(t)=\psi(t-i\Delta_{N}-\Delta_{N}/2-\ln L,\Delta_{N}/8)}\nonumber\\
&&\ \ \ \ \ \ \ \ \ \ \ \ \ \ \ \ \ \ \ \ \ \  \ \ \ \ \ \ \ \ \ \ \ \ \ \ \ \ \ \times\exp\left[-\frac{t-\ln L}{2}\right]
\end{eqnarray}
where $\Delta_{N}=(\ln H-\ln L)/N$ and $i\in\{0,1,...,N-1\}$.  This definition shifts and scales the location and size of the base wavelet so that an
integer number $N$ of them fit in the interval $[\ln L,\ln H]$ and they have most of their support in each width $\Delta_{N}$ subinterval of the base
interval $\Delta_{1}=\ln H - \ln L$.  Given this starting set of smooth functions, they are turned into E- and B-mode estimators by discretizing the
functions over the interval $[\ln L,\ln H]$ and then projecting out the $F_{+a}$ and $F_{+b}$ modes as described above.  The filters which result from
this process will be denoted as $F_{+Ni}$ and $F_{-Ni}$.  Figure \ref{fig:hatwavebasis} shows an example set of these filters for 100 correlation
function data points between 1 and 400 arcmin.

In order to gain intuition into how these B-mode estimators work, I show a simple, but somewhat contrived example \citep[however see e.g.,][]{fu2008,eifler2010}
of two different kinds of B-mode systematics in Figure \ref{fig:wvtest}.  The top row displays a smooth B-mode systematic in the $\xi_{-}$
correlation function, while the bottom row shows a B-mode systematic with much more fine scale variations.  These B-mode systematics were generated by 
setting $P_{B}(\ell)$ to be non-zero using simple Gaussian kernels, the form of which is uninteresting, and then computing the shear correlation functions using Equations~(\ref{eqn:xipell}) 
and (\ref{eqn:ximell}).  The contour plots show for each value of $N$ on the vertical axis, the amplitude of the B-mode statistics for each $\Delta_{N}$ sized interval across
the horizontal axis.  In this type of analysis, a B-mode signal with more fine scale variation has a different signature than the smoother B-mode
signal, with more B-mode signal detected in higher $N$ and thus smaller sized filters.  For data with shape noise, a similar contour plot can be made,
except that the deviation from zero measured in units of the standard deviation should be used to produce the color scaling.  The overall level of
B-mode contamination can then be assessed via a $\chi^{2}$ statistic computed over all of the different estimators indexed by $\{N,i\}$, accounting
for the correlations between the different estimators.  

\begin{figure}
\begin{center}
\includegraphics[scale=0.5]{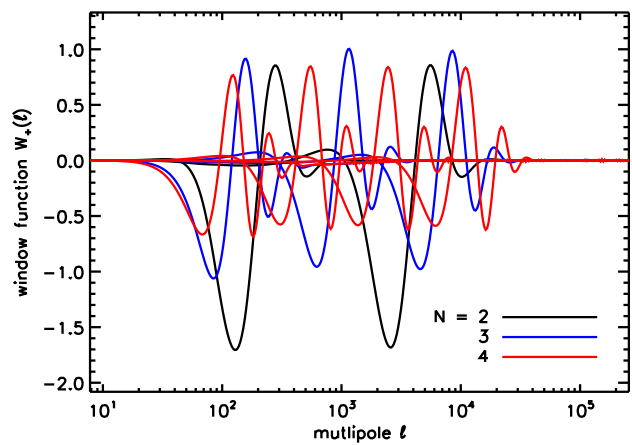}
\end{center}
\caption[]{The Fourier space filters $W_{+}(\ell)$ for the $N\in\{2,3,4\}$ wavelet-like statistics computed for $\theta\in[1,400]$ arcmin with 100 shear correlation 
function data points. The different colors show each of the filters.  The $W_{-}(\ell)$ filters are approximately three orders of magnitude smaller than the $W_{+}(\ell)$ 
filters.  The different filters in size (the overall width of the peaks and troughs) and location (the mean location in $\ell$ for each set of peaks and troughs) approximately 
cover all of $\ell$-space accessible given the angular range of the data.\label{fig:waveellwin}}
\end{figure}

Here I have chosen to layout the $F_{\pm Ni}$ filters in a simple geometric pattern.  However, when a B-mode signal crosses between two filters, the
significance drops.  Thus it might be more advantageous to consider sets of filters which shift and scale in size more continuously.  I leave this
generalization to future work.  Note however that the filters I have presented here are offset between different levels $N$ so that in practice, no B-mode signal
ever goes completely undetected.  This general point can be clearly illustrated by examining the Fourier-space filters $W_{+}(\ell)$ introduced above 
for the wavelet-like $F_{\pm Ni}$ statistics.  These filters for $N\in\{2,3,4\}$ are shown in Figure \ref{fig:waveellwin}.  The peaks and troughs for each $N$ correspond to a 
single location of the filter in real space, with $N$ peaks and troughs present for the $N$-indexed filters.  As the filters shift in real space over the shear correlation function 
range, they cover different, but overlapping ranges in Fourier space.  Thus by combining filters of different size and location, any B-mode signals in Fourier space 
will produce a nonzero amplitude in these filters.  

\subsection{Fisher Information Content and Full Mode Decomposition with COSEBI-like Statistics}\label{sec:fishinfo}
In this section I present the full ambiguous, E- and B-mode decompositions discussed in Section~\ref{sec:ebmodedec} using the COSEBI-like statistics defined above.  I also compute the Fisher 
information content of the approximate ambiguous, E- and B-mode subspaces for the parameters $\sigma_{8}$ and $\Omega_{m}$.  I use a basis composed of 50 correlation function bins 
logarithmically spaced from 1 to 400 arcmin. I find that the full basis composed of these different spaces completely spans 
the space of $2N$ data points, demonstrating that at least one such approximate decomposition exists.  

Due to the fact that this decomposition is only approximate and that the matrices $\mathrm{M}_{\pm}$ are not exactly orthogonal for the 
vectors $F_{\pm}$, the different subspaces will in general not be perfectly orthogonal to one another.  The degree of orthogonality of the ambiguous, E- and 
B-mode subspaces can be characterized by the maximum absolute normalized projection of a basis vector of any one of the 
spaces onto the basis vectors of the other spaces,
\begin{equation}
|\cos\theta| = \left|\frac{\vec{a}\cdot\vec{b}}{|\vec{a}||\vec{b}|}\right|
\end{equation}
where $\vec{a}$ and $\vec{b}$ are vectors of length $2N$ with the normal inner product definition.  This quantity is just the cosine of the minimum angle between the vectors 
of any of the subspaces with the other.  For subspaces which nearly share a basis vector, this maximum absolute normalized projection will be $\approx1$, but it cannot be greater 
than one or else two of the vectors in the basis would be linearly dependent.  For spaces which are mutually orthogonal, this maximum absolute normalized 
projection will be zero.  

For the set of basis vectors just computed, I find that the maximum projection between the E- and B-mode subspaces is $1.39\times10^{-1}$.  Similarly, the maximum projection 
between the E-mode and ambiguous subspaces is $1.51\times10^{-3}$ and for the B-mode and ambiguous subspaces is $5.52\times10^{-3}$.  Consistent with the argument made above, with equal 
area binning the E- and B-mode subspaces are more orthogonal with the maximum absolute normalized projection between them being $2.49\times10^{-2}$.  However in this case, the ambiguous modes 
are less orthogonal to the E- and B-mode subspaces with the maximum absolute normalized projections being $1.31\times10^{-1}$ and $3.19\times10^{-1}$ respectively.  The particularly poor orthogonality 
of the B-mode subspace to the ambiguous mode space in this case is reflected in the SVD of the matrix $\mathrm{M}_{+}\mathrm{M}_{-} - \mathrm{I}$ computed with equal area binning.

As mentioned above, the cost of E- and B-mode separation is that one must potentially throw away information in the ambiguous modes \citep[e.g.,][]{smith2006,schneider2010}.  Using the 
approximate mode decomposition described in this section, this point can be illustrated clearly and approached quantitatively.  I compute the Fisher information content of the ambiguous, E- 
and B-mode subspaces and combinations of them assuming Gaussian statistics and using the expressions presented above for the covariances of the two-point correlation functions 
and Fisher information matrices.  For the E- and B-mode statistics, I simply transform the covariance matrix of the two-point correlation functions, since the statistics are linear 
combinations of the two-point data. One can obtain similar results computing their covariance matrices directly from their Fourier space window functions 
and the expressions presented in Section~\ref{sec:ebcov}.  The covariance matrix of the ambiguous modes is computed by direct transformation as well.

The Fisher information content $f\equiv\sqrt{|F|}$ of these subspaces for the parameters $\sigma_{8}$ and $\Omega_{m}$ using the binning scheme and angular range for the basis constructed above 
and also assuming DES-like errors is $2.18\times10^{4}$, $3.49\times10^{4}$, and $4.49\times10^{-1}$ for the ambiguous, E- and B-mode subspaces respectively.  By comparison, the full 
Fisher information for the entire two-point function data set, $\{\xi_{+i},\xi_{-i}\}$ for $i\in[i,N]$, is $5.98\times10^{4}$. Note that these figures are not additive and so can only be 
qualitatively compared.  However the basic trends are clear.  The B-mode subspace has very little information on the parameters as expected.  Any residual information is either a numerical artifact or is 
due to E/B-mode mixing.  The E-mode subspace has most of the information, but the ambiguous modes carry a non-trivial fraction of the total information.  By retaining the ambiguous modes in the 
analysis with the E-modes, most of the information can be restored in comparison with the full two-point function with the E- plus ambiguous mode Fisher information being $5.75\times10^{4}$.  I have 
tested several other fiducial choices besides the COSEBI-like statistics for constructing the full mode decomposition of the correlation function data and found similar conclusions regarding the relative 
Fisher information content of the various subspaces.  Similar conclusions hold for LSST-like surveys as well.  While these Fisher information calculations are not particularly realistic as survey projections, 
they illustrate nicely the cost of E/B-mode separation in cosmic shear.

Finally, these Fisher information calculations can also illustrate the data compression properties of the COSEBI-like statistics.  In particular, for the DES-like survey, the 
Fisher information in the first eight E-mode statistics is $3.46\times10^{4}$ as opposed to $3.49\times10^{4}$ for all of the E-mode statistics.  Similarly, 
for the LSST-like survey using 100 correlation function bins in the same angular range as the DES-like survey, the first ten E-mode statistics have an information content of 
$6.489\times10^{5}$ as opposed to $6.496\times10^{5}$ for all of the E-mode statistics.  These results are in qualitative agreement with those of \citet{schneider2010}.

\section{Conclusions}\label{sec:conc}
In this work, I have demonstrated that the use of two-point E/B-mode estimators with binned shear correlation function data will generally result in non-trivial E/B-mode 
mixing.  I computed the amount of this mixing and provided new E/B-mode estimators which minimize the unwanted mode mixing. I also gave practical 
recipes for building and using these estimators with binned cosmic shear data.  Using these optimal estimators, I demonstrated that approximate decompositions of the vector space of binned 
correlation function points into ambiguous, E- and B-modes do exist.  Using one of these decompositions, I found that the ambiguous mode subspace contains a non-trivial amount of information on typical 
cosmological parameters.  I also gave two example applications of these new estimators to generic problems in cosmic shear data analysis, data compression (Section~\ref{sec:cosebi}) and B-mode 
localization with wavelets (Section~\ref{sec:wvstats}).

The estimators presented here have several nice features adapted to practical cosmic shear data analysis.
\begin{itemize}
\item They are linear combinations of the binned shear correlation function data, defined in Equation~(\ref{eqn:xpmlinstat}), and thus treat the binning explicitly.  This property makes their computation and 
also error propagation/analysis with them trivial once the shear correlation function and its errors are known.
\item The level of E/B-mode mixing due to the binning can be computed exactly, up to the knowledge of the binning window functions, for these estimators 
using Equations~(\ref{eqn:xipmell}) and (\ref{eqn:wpmdef}).  In the limit of small bins, which is needed to suppress the E/B-mode mixing, the binning window functions are expected to be close to the 
geometric approximation used throughout this work.
\item They give quantitative criteria in terms of the E/B-mode mixing by which to decide the number of shear correlation 
function bins to use, demonstrated in Figure~\ref{fig:cosebiss2n} for the COSEBI-like estimators.
\item The design of new E/B-mode estimators with specific purposes using the formalism presented in this work reduces to simple linear algebraic manipulations, presented in Section~\ref{sec:buildEBest}.
\end{itemize}

The optimal statistics presented in this work are well-suited to blinded or closed-box, high-precision cosmic shear data analyses.  For example, before any shear correlation functions are 
computed from the data, one can estimate given the expected statistical accuracy of the data, the exact amount of E/B-mode mixing for a given binning scheme and set of estimators.  
One can then choose a fiducial binning scheme and estimator choice that properly minimizes the E/B-mode mixing and retains all of the cosmological information.  Then these choices can be fixed 
throughout the data analysis process in order to avoid any observer biases in detecting or assessing B-mode contamination arising from how exactly the E/B-mode separation was done. Additionally, in a 
blinded or close-box analysis one might not look at plots of the shear correlation function data until the entire analysis is complete.
Unfortunately in this case, one might miss crucial information about potential systematics in the \textit{shape} of the shear correlation function data. The
wavelet-like B-mode estimators presented in this work can be used as a substitute for and quantitative measure of the information gained by looking at the
shape of the shear correlation function data.  Additionally, they can be applied in automated way to large data sets in order to pinpoint areas of potential systematic 
contamination for further investigation.  Importantly, because these estimators have a large degree of spatial locality, they can potentially provide crucial information on where 
any B-mode contamination is coming from and not just indicate its existence.  

The ability to easily and quickly design E/B-mode estimators tailored to a specific purpose can potentially be very useful in practice as well.  For
instance, the problem of computing an E/B-mode statistic which maximizes the signal-to-noise or cosmological information content, as considered by
\citet{fu2010}, could now be reformulated with the linear estimators presented here.  Additionally, one could build estimators which are along the
normal modes of the correlation function data computed from the correlation function covariance matrix.  Also, one could attempt to build direct estimators
for the E/B-mode correlation functions \citep{crittenden2002,schneider2002,schneider2010} over a finite interval.  Alternatively, one could attempt to build
filters which are localized in Fourier space in order to exclude certain wave modes, for example to mitigate the effects of uncertainty in the matter
power spectrum at small scales \citep[see e.g.,][]{huterer2005b}, or to be sensitive to B-modes from only a range of $\ell$. As simple linear combinations of 
cosmic shear data points, the estimators presented in this work are well-suited to these applications and to practical cosmic shear data analysis in general.  

\section*{Acknowledgments}

I thank Tim Eifler, Eduardo Rozo, Scott Dodelson, Bhuvnesh Jain, Aaron Roodman, and the DES Weak Lensing Working Group for the many 
discussions which inspired aspects of this work. This work was supported by the Kavli Institute for Cosmological Physics at the University of 
Chicago through grants NSF PHY-0114422 and NSF PHY-0551142 and an endowment from the Kavli Foundation and its founder Fred Kavli.
This work made extensive use of the NASA Astrophysics Data System and {\tt arXiv.org} preprint server.

\appendix

\section{Pair-wise Shear Correlation Function Estimators and Finite Bin Widths}\label{app:srcclust}
I use the formalism presented in Appendix A of \citet{schmidt2009a} to derive the leading order bin weighting terms in the pair-wise estimator for
the shear correlation functions.  This estimator is \citep[e.g.,][]{schneider2002b}
\begin{equation}
\widehat{\xi}_{\pm k} = \frac{\sum_{ij}w_{i}w_{j}W_{\theta_{k}}(x_{i},x_{j})(\epsilon_{it}\epsilon_{jt} \pm \epsilon_{i\times}\epsilon_{j\times})}{\sum_{ij}w_{i}w_{j}W_{\theta_{k}}(x_{i},x_{j})}
\end{equation}
where $W_{\theta_{k}}(x_{1},x_{2})$ is the binning function which is unity inside the bin and zero outside with the bin centered at $\theta_{k}$ with some bin width $\Delta\theta$, 
$\epsilon_{it,i\times}$ is the component of the galaxy shape parallel or crossed with respect to the great circle connecting the two galaxies, and the $w_{i,j}$ are weights applied to each galaxy.  
Below I include the weights $w_{i,j}$ in the survey window function.  Starting with Equation A15 of \citet{schmidt2009a} and assuming that the source galaxy density field is uncorrelated
with the shear field and also neglecting lensing bias/reduced shear effects, the expectation value of this estimator can be written as
\begin{eqnarray}
\lefteqn{\left<\hat{\xi}_{ab}\right> = \frac{1}{\Omega_{B}}\int d^{2}x_{1}\int d^{2}x_{2}\, W_{\theta}(x_{1},x_{2})S(x_{1})S(x_{2})}\nonumber\\
&&\times\,\xi_{ab}(|x_{1}-x_{2}|)\bigg[1 +\xi_{gg}(|x_{1}-x_{2}|) - \nonumber\\
&&\ \ \ \ \ \ \ \ \ \ \ \ \ \frac{1}{\Omega_{B}}\int d^{2}x_{3}\int d^{2}x_{4}\, W_{\theta}(x_{3},x_{4})S(x_{1})S(x_{2})\nonumber\\
&&\ \ \ \ \ \ \ \ \ \ \ \ \ \ \ \ \ \ \ \  \ \ \ \ \ \ \ \ \ \ \ \ \ \ \ \ \ \ \times\,\xi_{gg}(|x_{3}-x_{4}|)\bigg]
\end{eqnarray}
where $\xi_{ab}$ is the shear correlation function for $a,b\in\{t,\times\}$, $\Omega_{B} = \int d^{2}x_{1}\int d^{2}x_{2}\, W_{\theta}(x_{1},x_{2})S(x_{1})S(x_{2})$ is
the survey averaged bin area, and $\xi_{gg}$ is the source galaxy angular correlation function.  $S(x_{1})$ is the survey window function including the weights $w_{i,j}$. 
The second term in the brackets arises directly from the weighting over the bin by the sampling density of the 
source galaxies and the third term in the brackets is the first non-trivial term in the power-series expansion of the denominator of the estimator (i.e., 
the total number of observed galaxies in the bin).  The last two terms in this integral do not exactly cancel as stated in \citet{schmidt2009a} 
because $\xi_{gg}$ is not exactly equal to its average over the bin for all $\theta$ in the bin.  In addition to this source clustering weighting term, the survey window function 
contributes an additional weight over the bin.  The form of this weighting function is highly non-trivial, but for small 
enough bins, this weighting should be negligible.  However, quantitative results describing how small the bins need to be in order to suppress this weighting require 
detailed survey simulations, which are beyond the scope of this work.  Thus for simplicity I neglect the survey window function weights.  
A simple estimate of the magnitude of the effect of source galaxy 
clustering using typical galaxy angular correlation functions \citep[e.g.,][]{connolly2002} shows that this effect (i.e., the difference of the last two terms in the bracket above) 
is $\la0.005$ when using at least five bins per decade.  To get this number I assumed the galaxy-galaxy angular correlation function was a power law over the angular range of
$[1,400]$ arcmin with $\xi_{gg}=0.0045\times(\theta/1\ \mathrm{deg})^{-0.7}$, consistent with the results of \citet{connolly2002}. 

\section{Optimal E/B-mode Estimators with Geometric Bin Weighting Functions}\label{app:estgeomwgt}
In this Appendix, I give the exact form of the constraint directions, $F_{+a}$ and $F_{+b}$, and the matrix to compute the $F_{-i}$ from the $F_{+i}$ under the assumption of 
geometric bin weightings.  Consider $N$ bins in angle $\theta$ from $L$ to $H$ and let $[L_{i},H_{i}]$ be the angular range of the $i$th bin.  Also, assume the bins are non-overlapping. 
Then using the geometric bin weighting function, $W_{i}(\theta)=2\theta/(H_{i}^{2}-L_{i}^{2})$, I get for the constraint direction vectors 
(see Section~\ref{sec:buildEBest})
\begin{eqnarray}
F_{+a} &=& \left(1,1,1,...,1\right)_{N}\\
F_{+b} &=& \left(\frac{H_{1}^{4}-L_{1}^{4}}{2(H_{1}^{2}-L_{1}^{2})},\frac{H_{2}^{4}-L_{2}^{4}}{2(H_{2}^{2}-L_{2}^{2})},\frac{H_{3}^{4}-L_{3}^{4}}{2(H_{3}^{2}-L_{3}^{2})},\right.\nonumber\\
&&\left.\ \ \ \ \ \ \ \ \ \ \ \ \ \ \  \ \ \ \ \ \ \ \ \ \ \ \  ...,\frac{H_{N}^{4}-L_{N}^{4}}{2(H_{N}^{2}-L_{N}^{2})}\right)_{N}\ .
\end{eqnarray}
The matrix to compute the $F_{-k}$ in terms of the $F_{+i}$ from Equation~(\ref{eqn:fmdef}) is 
\begin{eqnarray}
\lefteqn{(\mathrm{M}_{+})_{ki} = \delta_{ki} + \frac{2}{H_{i}^{2}-L_{i}^{2}}}&&\nonumber\\
&&\times\left\{
\begin{array}{ll}
2(H_{i}^{2}-L_{i}^{2})\ln\left(H_{k}/L_{k}\right)&\\
 \ \ \ \ \ \ + \frac{3}{2}\left(H_{i}^{4} - L_{i}^{4}\right)\left(\frac{1}{H_{k}^{2}} - \frac{1}{L_{k}^{2}}\right) & \mbox{if $i < k$}\\
-\frac{1}{2}\left(H_{k}^{2}-L_{k}^{2}\right)  - 2L_{i}^{2}\ln\left(H_{k}/L_{k}\right)&\\
 \ \ \ \ \ \  -\frac{3}{2}L_{i}^{4}\left(\frac{1}{H_{k}^{2}} - \frac{1}{L_{k}^{2}}\right) & \mbox{if $i = k$}\\
0 & \mbox{if $i > k$}
\end{array}
\right.
\end{eqnarray}
where $i$ and $k$ run over 1 to $N$.  Then the $F_{-k}$ are computed as
\begin{eqnarray}
F_{-k} = \sum_{i=1}^{N}(\mathrm{M})_{ki}\,F_{+i}\ .
\end{eqnarray}

\section{Optimal E/B-mode Estimators with $F_{-}$ Fixed}\label{app:fminusest}
One can easily define optimal E/B-mode estimators with $F_{-}$ fixed instead of $F_{+}$ fixed.  In this case one fixes the $F_{+}$ by minimizing 
\begin{equation}
0=\frac{\partial}{\partial F_{+k}}\left[\int_{0}^{\infty}d\ell\,\ell|W_{-}(\ell)|^{2}\right]\ .
\end{equation}
The solution to this equation is (c.f. Equation~\ref{eqn:fmdef})
\begin{eqnarray}\label{eqn:fm2fpmat}
\lefteqn{F_{+k} = F_{-k}  + \left(\int_{L_{k}}^{H_{k}}d\theta\frac{W_{k}^{2}(\theta)}{\theta}\right)^{-1}\sum_{i}F_{-i}}&&\nonumber\\
&&\times\int_{L_{i}}^{H_{i}}\int_{L_{k}}^{H_{k}}d\theta\,d\phi\,W_{i}(\theta)\,W_{k}(\phi)\nonumber\\
&&\ \ \ \ \  \ \ \ \ \ \ \ \ \ \ \ \ \ \ \ \ \ \ \ \ \ \ \times\left(\frac{4}{\theta^{2}} - \frac{12\phi^{2}}{\theta^{4}}\right)H(\theta-\phi)\,.
\end{eqnarray}
In this case the constraint directions can be derived by similar arguments to the those in Section~\ref{sec:buildEBest} and are
\begin{eqnarray}
\lefteqn{F_{-a} = \left(\int_{L_{1}}^{H_{1}}d\theta\,\frac{W_{1}(\theta)}{\theta^2},\int_{L_{2}}^{H_{2}}d\theta\,\frac{W_{2}(\theta)}{\theta^2},\right.}\nonumber\\
&&\ \ \ \ \ \ \ \ \ \ \ \ \ \ \ \ \ \ \ \ \ \ \ \ \ \ \ \ \ \ \ \ \ \left. ...,\int_{L_{N}}^{H_{N}}d\theta\,\frac{W_{N}(\theta)}{\theta^2}\right)_{N}\\
\lefteqn{F_{-b} = \left(\int_{L_{1}}^{H_{1}}d\theta\,\frac{W_{1}(\theta)}{\theta^4},\int_{L_{2}}^{H_{2}}d\theta\,\frac{W_{2}(\theta)}{\theta^4},\right.}\nonumber\\
&&\ \ \ \ \ \ \ \ \ \ \ \ \ \ \ \ \ \ \ \ \ \ \ \ \ \ \ \ \ \ \ \ \ \left. ...,\int_{L_{N}}^{H_{N}}d\theta\,\frac{W_{N}(\theta)}{\theta^4}\right)_{N}\ 
\end{eqnarray}
and Equation~(\ref{eqn:fm2fpmat}) again defines a matrix which is used to compute the $F_{+}$ in terms of the $F_{-}$.
Finally, for the geometric bin weighting functions the constraint directions and matrix relating the $F_{+}$ to $F_{-}$ are
\begin{eqnarray}
\lefteqn{F_{-a} = \left(\frac{2\log[H_{1}/L_{1}]}{H_{1}^{2}-L_{1}^{2}},\frac{2\log[H_{2}/L_{2}]}{H_{2}^{2}-L_{2}^{2}},\right.}\nonumber\\
&&\ \ \ \ \ \ \ \ \ \ \ \ \ \ \ \ \ \ \ \ \ \ \ \ \ \ \ \ \ \ \ \ \ \ \left. ...,\frac{2\log[H_{N}/L_{N}]}{H_{N}^{2}-L_{N}^{2}}\right)_{N}\\
\lefteqn{F_{-b} = \left(\frac{1}{H_{1}^{2}-L_{1}^{2}}\left(\frac{1}{L_{1}^2} - \frac{1}{H_{1}^2}\right),\frac{1}{H_{2}^{2}-L_{2}^{2}}\left(\frac{1}{L_{2}^2} - \frac{1}{H_{2}^2}\right),\right.}\nonumber\\
&&\ \ \ \ \ \ \ \ \ \ \ \ \ \ \ \ \ \ \ \ \left. ...,\frac{1}{H_{N}^{2}-L_{N}^{2}}\left(\frac{1}{L_{N}^2} - \frac{1}{H_{N}^2}\right)\right)_{N}
\end{eqnarray}
and
\begin{eqnarray}
\lefteqn{(\mathrm{M}_{-})_{ki} = \delta_{ki} + \frac{2}{H_{i}^{2}-L_{i}^{2}}}&&\nonumber\\
&&\times\left\{
\begin{array}{ll}
0 & \mbox{if $i < k$}\\
\frac{1}{2}\Big[-H_{k}^{2}
                         + L_{k}^{2}\,(4 - 3L_{k}^{2}/H_{i}^{2} &\\
                         \ \ \ \ \ \ \ \ - 4\ln(H_{i}/L_{k}))\Big] & \mbox{if $i = k$}\\
2(H_{k}^{2}-L_{k}^{2})\ln\left(H_{i}/L_{i}\right) &\\
\ \ \ \ \ \ \ \ + \frac{3}{2}\left(H_{k}^{4} - L_{k}^{4}\right)\left(\frac{1}{H_{i}^{2}} - \frac{1}{L_{i}^{2}}\right) & \mbox{if $i > k$}\\
\end{array}
\right.
\end{eqnarray}
where $i$ and $k$ run over 1 to $N$.  Then the $F_{+k}$ are computed as
\begin{eqnarray}
F_{+k} = \sum_{i=1}^{N}(\mathrm{M})_{ki}\,F_{-i}\ .
\end{eqnarray}


\bsp

\label{lastpage}

\end{document}